\documentclass[reprint, superscriptaddress, amsmath,amssymb, aps, pra]{revtex4-1}

\usepackage{color}

\usepackage{graphicx}% Include figure files
\usepackage{dcolumn}% Align table columns on decimal point
\usepackage{bm}% bold math
\usepackage{siunitx}
\usepackage{upgreek}
\begin{document}

\preprint{APS/123-QED}

%\title{Sphere in a Bottle: The Dark Side of Optical Trapping}

\title{Optical Trapping in a Dark Focus}
%Optical trapping 
%Sphere in a Bottle:  Optical Trapping in the Dark
\author{B. Melo}
\email{brunomelo@aluno.puc-rio.br}
\affiliation{Departamento de F\'{i}sica, Pontif\'{i}cia Universidade Cat\'{o}lica do Rio de Janeiro,  22451-900 Rio de Janeiro, RJ, Brazil}
\author{I. Brand\~{a}o}
\email{igorbrandao@aluno.puc-rio.br}
\affiliation{Departamento de F\'{i}sica, Pontif\'{i}cia Universidade Cat\'{o}lica do Rio de Janeiro,  22451-900 Rio de Janeiro, RJ, Brazil}
\author{B. Pinheiro da Silva}
\email{braianps@gmail.com}
\affiliation{Instituto de F\'{i}sica, Universidade Federal Fluminense, 24210-346 Niter\'{o}i, RJ, Brazil}
\author{R. B. Rodrigues}
\email{rafaelbellasrodrigues@gmail.com}
\affiliation{Instituto de F\'{i}sica, Universidade Federal Fluminense, 24210-346 Niter\'{o}i, RJ, Brazil}
\author{A. Z. Khoury}
\email{azkhoury@id.uff.br}
\affiliation{Instituto de F\'{i}sica, Universidade Federal Fluminense, 24210-346 Niter\'{o}i, RJ, Brazil}
\author{T. Guerreiro}
\email{barbosa@puc-rio.br}
\affiliation{Departamento de F\'{i}sica, Pontif\'{i}cia Universidade Cat\'{o}lica do Rio de Janeiro,  22451-900 Rio de Janeiro, RJ, Brazil}
%\author{L. Defaveri}
%\affiliation{Departamento de Física, Pontifícia Universidade Católica do Rio de Janeiro,  22451-900 Rio de Janeiro, RJ, Brazil}

\date{\today}

\begin{abstract}
The superposition of a Gaussian mode and a Laguerre-Gauss mode with $\ell=0,p\neq0$ generates the so-called bottle beam: a dark focus surrounded by a bright region. In this paper, we theoretically explore the use of bottle beams as an optical trap for dielectric spheres with a refractive index smaller than that of their surrounding medium. 
The forces acting on a small particle are derived within the dipole approximation and used to simulate the Brownian motion of the particle in the trap. The intermediate regime of particle size is studied numerically and it is found that stable trapping of larger dielectric particles is also possible. 
Based on the results of the intermediate regime analysis, an experiment aimed at trapping living organisms in the dark focus of a bottle beam is proposed. 
%\begin{description}
%\end{description}
\end{abstract}

\maketitle

%\tableofcontents

\section{\label{sec:introduction} Introduction}

%When a dielectric particle is immersed in a medium, and the refractive index of the particle is \textit{higher} than that of its surroundings, an optical force generated by an incident laser beam pulls the particle to a region of high intensity of light. This technique, introduced by Arthur Ashkin in 1986 \cite{Ashkin1986} and known today as optical tweezing, finds applications in a large number of fields ranging from biology \cite{Fazal2011, Nussenzveig2017, S.Araujo2019} to fundamental physics \cite{Monteiro2020, Ether2015, Arvanitaki2013, Geraci2010, Moore2014}. In standard optical tweezers, Gaussian beams are used to create the trapping focus. To a good approximation, the trap can be described as a three dimensional quadratic potential. 

Tightly focused laser beams can be used to exert forces upon dielectric particles. If the particle's refractive index is \textit{larger} than that of its surroundings, the laser pulls it to regions of higher intensity of light. This technique, introduced by Arthur Ashkin in 1986 \cite{Ashkin1986} and known today as optical tweezing, allows one to hold and manipulate very tiny objects and finds applications in a large number of fields ranging from biology \cite{Fazal2011, Nussenzveig2017, S.Araujo2019, Pontes2013} to fundamental physics \cite{Monteiro2020, Ether2015, Arvanitaki2013, Geraci2010, Moore2014}. In standard optical tweezers, Gaussian beams are used to create the trapping focus. To a good approximation, the trap can be described as a three dimensional quadratic potential.  

%A laser beam incident on a dielectric particle of refractive index \textit{higher} than that of its surroundings pulls the particle to the region of higher intensity of light.  This technique, introduced by Arthur Ashkin in 1986 \cite{Ashkin1986} and known today as optical tweezing, allows to hold and manipulate very tiny particles and finds applications in a large number of fields ranging from biology \cite{Fazal2011, Nussenzveig2017, S.Araujo2019} to fundamental physics \cite{Monteiro2020, Ether2015, Arvanitaki2013, Geraci2010, Moore2014}. In standard optical tweezers, Gaussian beams are used to create the trapping focus, generating, to a good approximation, the trap can be described as a three dimensional quadratic potential. 

Notably it was also pointed out by Ashkin that air droplets immersed in water were pushed away from the Gaussian focus \cite{Ashkin1970}. This is a consequence of the fact that when the refractive index of the particle is \textit{smaller} than that of its surroundings, the particle is repelled from the region of high intensity. One can then envision an \textit{inverted} optical trap, in which an engineered beam of light has a high-intensity boundary and a dark focus. A particle with the appropriate refractive index will be trapped within the dark focus by the absence of light \cite{Ahluwalia2006}. 
We refer to this type of beam in general as \textit{bottle beams}. A bottle beam is one example in a myriad of engineered optical traps aimed at different purposes such as circular Airy beams \cite{Lu2019, Cheng2010, Jiang2013}, Bessel beams \cite{Arlt2001}, radially polarized beams \cite{Yan2007, Shu2013}, frozen waves \cite{Suarez2020} and many others \cite{Zhao2007, Zhao2009, Zhao2009a,Zhang2015, Zhao2011, Zhan2003}. 

Several techniques can be employed to create bottle beams, such as the generation of Bessel beams using axicons \cite{Wei2005, Lin2007, Du2014}, the interference of Gaussian beams of different waists \cite{Isenhower2009} and the superposition of different modes \cite{Silva2020, Ahluwalia2004, Xu2010, Zhang2011} created using Spatial Light Modulators \cite{Matsumoto2008, Ohtake2007, Rhodes2006, Ando2008}. Here, we focus on the bottle beam created by the superposition of a Gaussian beam and a Laguerre-Gauss beam with $\ell=0, p\neq0$ and a relative phase of $\pi$ presented in \cite{Arlt2000} and study the optical forces it exerts upon low refractive index particles.

Because optical trapping can be applied to particles in a wide size range \cite{Hansen2005, Alinezhad2019}, we analyse both the cases of small Rayleigh particles and of larger micron-sized particles. In the former, the optical forces and potential are derived from the dipole approximation and thoroughly analysed under different assumptions, which are verified by simulating the motion of the trapped particle in a viscous medium. In the latter, generalized Lorenz–Mie theory is employed to calculate the forces caused by the beam, with the aid of the tools introduced in \cite{Nieminen2007}. Constraints on the numerical aperture, particle size and relative refractive index are found.

Understanding particle dynamics under the influence of a bottle beam can lead to striking applications. Notably, the bottle is an interesting tool for trapping experiments requiring little or no light scattering upon the trapped object.
This is of particular interest in biology, where trapping a living cell or organelles within the cell without the constant influence of laser light might be crucial to reveal mechanical properties of the organism without excessive heating and laser interference \cite{Liu1995, Peterman2003, BlazquezCastro2019}. We thus propose a set of experimental parameters that could be used to trap living organisms in the dark focus of a bottle beam.

\section{The dipole approximation}
\label{sec:dipole}

We begin by investigating the optical forces acting on a Rayleigh particle with a refractive index lower than its surrounding medium  under the dipole approximation. 
This is valid when the radius of the trapped particle is much smaller than the wavelength of the trapping laser $(R\lesssim\lambda/10)$ \cite{Li2013}. %how smaller, R < \lambda / 10?%

%We begin by investigating the optical forces acting on a low refractive index particle in the Rayleigh regime using the dipole approximation, which is valid when the radius of the trapped particle is much smaller than that of the laser wavelength.

\subsection{The optical bottle beam}

To generate a dark focus surrounded by a bright region we superpose a Laguerre-Gauss beam with $\ell=p=0$ - a Gaussian beam - and a Laguerre-Gauss beam with $\ell=0, p\neq0$ and a relative phase of $\pi$. The electric field magnitude of a Laguerre-Gauss beam is
%  \begin{eqnarray}
 % \label{eq:LGbeam}
  %    \hspace*{-2em}&\,&E^{LG}_{\ell,p}(\rho,\phi,z) =\sqrt{\frac{4P_0}{c\epsilon\pi\omega(z)^2}}\sqrt{\frac{p!}{(\vert \ell \vert+p)!}}\times\nonumber\\&\times&\exp\left[-\frac{\rho^2}{\omega(z)^2}+ik_mz-i\zeta(z)+ik_m\frac{\rho^2}{2R(z)} \right]\times\nonumber\\&\times&\left( \frac{\sqrt{2}\rho}{\omega(z)} \right)^{\vert\ell\vert}\hspace{-2mm}L^\ell_p\left( \frac{2\rho^2}{\omega(z)^2} \right)\exp[-i(2p+\vert\ell\vert)\zeta(z)+i\ell\phi],
%  \end{eqnarray}
\begin{eqnarray}
  \label{eq:LGbeam2}
      \hspace*{-2em}&\,&E^{LG}_{\ell,p}(\rho,\phi,z) =\sqrt{\frac{4P_0}{c\epsilon\pi\omega(z)^2}}\sqrt{\frac{p!}{(\vert \ell \vert+p)!}}\times\nonumber\\&&\left( \frac{\sqrt{2}\rho}{\omega(z)} \right)^{\vert\ell\vert}L^{\vert\ell\vert}_p\left( \frac{2\rho^2}{\omega(z)^2} \right)\exp\left[-\frac{\rho^2}{\omega(z)^2}\right]\times\nonumber\\&&\exp[ik_mz+ik_m\frac{\rho^2}{2R(z)}-i\zeta(z)+i\ell\phi],
\end{eqnarray}
where $c$ is the speed of light, $\epsilon$ is the medium's permittivity, $P_0$ is the laser power, $k_m$ is the wavenumber in the medium and $\omega(z)$, $R(z)$, $\zeta(z)$ and $L^{\vert\ell\vert}_p$ %(pela definição do Arfken $\ell>-1$, então devemos usar $\vert\ell\vert$) 
are the beam width, the wavefront radius, the Gouy phase and the Associated Laguerre polynomial. These quantities are respectively given by
\begin{eqnarray}
\omega(z)&=&\omega_0\sqrt{1+\frac{z^2}{z^2_R}};\\
  R(z) &=& z\left(1+\frac{z_R^2}{z^2} \right);\\
\zeta(z)&=&(2p+\vert\ell\vert+1)\arctan \frac{z}{z_R};\\
%  \zeta(z)&=&\arctan\frac{z}{z_R};\\
    L_p^{\vert\ell\vert}(x) &=& \sum^p_{i=0}\frac{1}{i!}\binom{p+\vert\ell\vert}{p-i}(-x)^i
    %L_p^\ell(x) &=& \sum^p_{i=0}\frac{1}{i!}\binom{p+\ell}{p-i}(-x)^i
\end{eqnarray}
%with $z_R = \pi \omega_0^2/\lambda_m$ the Rayleigh range, $\omega_0$ the beam's waist and $\lambda_0$ the wavelength in vacuum. Throughout this paper we always consider the electric fields to be linearly polarized. The intensity of the bottle beam can be written as %throughout space can be written as
\noindent where the Rayleigh range ($z_R$) and the beam waist ($\omega_0$) are defined as
\begin{eqnarray}
\omega_0=\frac{\lambda_0}{\pi \textrm{NA}}\,,\quad z_R = \frac{n_{m}\lambda_0}{\pi \textrm{NA}^2} \label{eq:waist_and_Rayleigh_range}
\end{eqnarray}

\noindent with $\lambda_0$ the wavelength in vacuum, $n_{m}$ the medium refractive index and $\textrm{NA}$ the numerical aperture. Throughout this work we will consider linearly polarized electric fields only.

\begin{figure}[t]
    \centering
    \includegraphics[width=\linewidth]{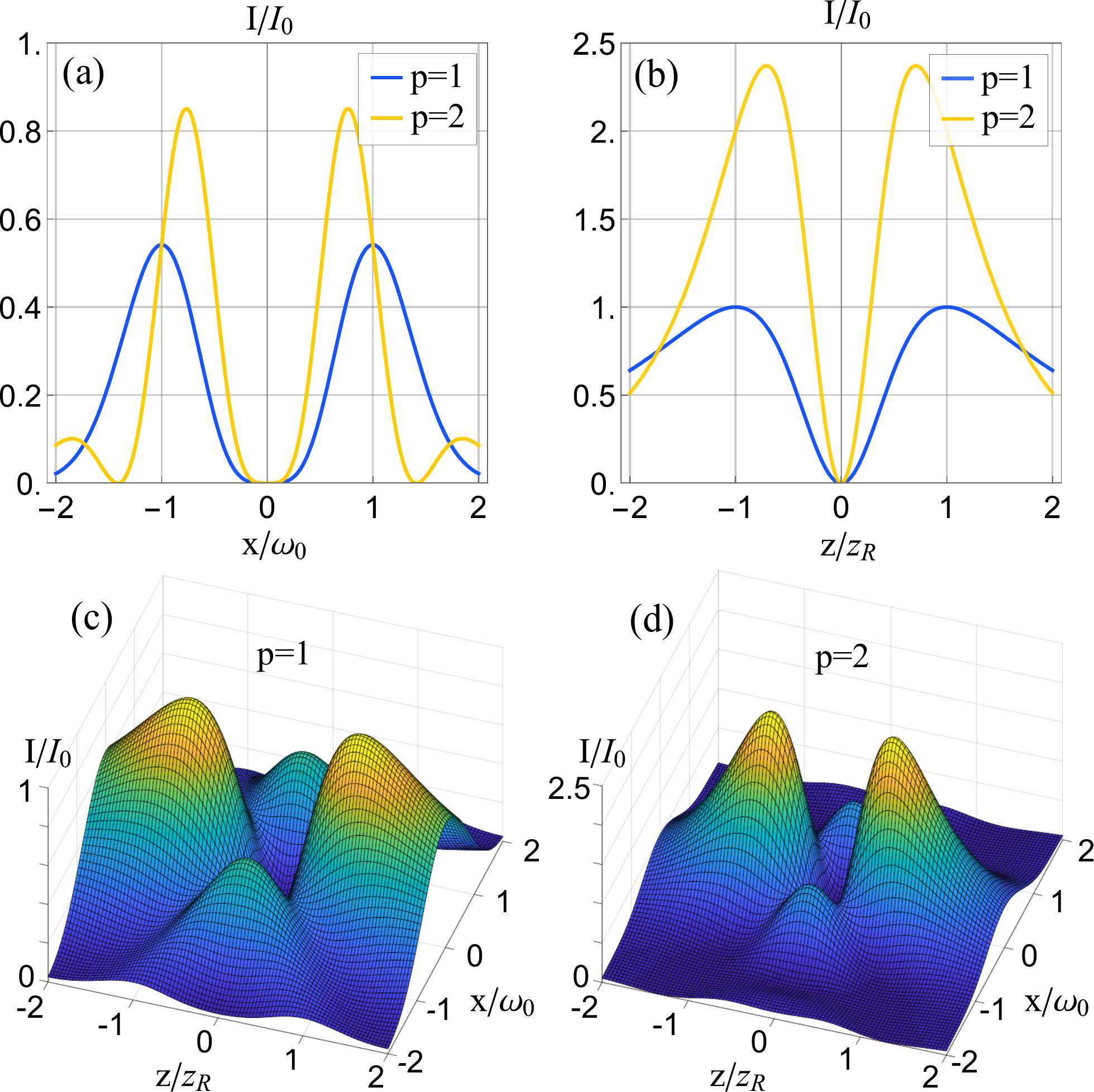}
    \caption{Intensity in the (a) radial and (b) axial directions for bottle beams with $p=1$ and $p=2$. Intensity landscape in the $xz$ plane for bottle beams with (c) $p=1$ and (d) $p=2$. Due to the normalization of $x$, $z$ and $I$, these plots depend only on $ p $, and are independent from the remaining beam parameters.}
    \label{fig:intensity_examples}
\end{figure}

The intensity of the bottle beam reads %throughout space can be written as
\begin{eqnarray}
\label{eq:exact_intensity}
&\,&I_p(\rho,z)=I_0\frac{\omega_0^2}{\omega(z)^2}\exp\left[-\frac{2 \rho^2}{\omega(z)^2}\right]\times\nonumber\\
&&\bigg[1-\hspace{-0.5mm}2\cos\left(2p\arctan\frac{z}{z_R}\right) L^0_p\left(\frac{2\rho^2}{\omega(z)^2}\right)+\nonumber\\&&L^0_p\left(\frac{2\rho^2}{\omega(z)^2}\right)^2 \bigg]
\end{eqnarray}
where $I_0=2P_0/\pi\omega_0^2$ is the intensity at the origin of the Gaussian beam. Figures \ref{fig:intensity_examples}(a) and \ref{fig:intensity_examples}(b) shows the intensity as a function of the transverse coordinate $x$ and the longitudinal coordinate $z$. The potential landscape in the $xz$ plane is shown in Figures \ref{fig:intensity_examples}(c) for the cases $p=1$, and \ref{fig:intensity_examples}(d) $p=2$. A dielectric particle with the appropriate refractive index placed at the origin would be trapped in the dark focus, since it would be repelled in all directions by the surrounding regions of higher electromagnetic intensity. 

\subsection{Dimensions of the bottle}
We can define the width $W$ (height $H$) of the bottle as the distance between the two intensity maxima surrounding the dark region along the $x$ axis ($z$ axis). These values can be found by solving
\begin{eqnarray}
\label{eq:width}
dI_p(x,0,0)/dx\vert_{x=W/2}=0\, ,\\
\label{eq:height}
dI_p(z,0,0)/dz\vert_{z=H/2}=0\, .
\end{eqnarray}

The above equations admit analytical solutions for small $p$, yielding $W=2\omega_0,H=2z_R$ for $p=1$ and $W=2\sqrt{2-\sqrt{2}}\omega_0,H=\sqrt{2}z_R$ for $p=2$. 

To gain insight into $H$ and $W$ it is useful to make the change of variables $\rho/\omega_0\rightarrow\rho', z/z_R\rightarrow z'$ in the intensity given in Eq.(\ref{eq:exact_intensity}). The function $I_p(\rho',z')$ has no explicit dependence on any of the beam's parameters other than $p$ and $I_0$, with its associated re-scaled width $W'$ and height $H'$. The pre-factor $I_0$ does not alter the distance between maxima along the $x'$ and $z'$ axis, meaning that $W' = W'(p)$ and $H' = H'(p)$ depend only on $ p $. 
%, so when we calculate the width and the height in this new coordinate system, we find solutions $W'(p)$ and $H'(p)$ that only depend on $p$. 
Going back to the original variables we find that $W=\omega_0W'(p)$ and $H=z_RH'(p)$. 
%For small $p$, Eqs. (\ref{eq:width}) and (\ref{eq:height}) admit analytical solutions. For $p=1$: $W=2\omega_0$, $H=2z_R$. For $p=2$: $W=2\sqrt{2-\sqrt{2}}\omega_0$, $H=\sqrt{2}z_R$. 
%The above equations admit analytical solutions for small $p$, yielding $W=2\omega_0,H=2z_R$ for $p=1$, and $W=2\sqrt{2-\sqrt{2}}\omega_0,H=\sqrt{2}z_R$ for $p=2$. 
%Alternatively, we can make the change of variables $\rho/\omega_0\rightarrow\rho', z/z_R\rightarrow z'$ in Eq.(\ref{eq:exact_intensity}), which gives $I_p(\rho',z')$ with no explicit dependence on any of the beam's parameter other than $p$ and the prefactor $I_0$. This prefactor doesn't alter the distance between the maxima along the $x'$ or $z'$ axis, so when we calculate the width and the height in this new coordinate system, we find solutions $W'(p)$ and $H'(p)$ that only depend on $p$. Going back to the initial coordinate system, we find that $W=\omega_0W'(p)$ and $H=z_RH'(p)$.
From Eq. (\ref{eq:waist_and_Rayleigh_range}), we observe that the width of the bottle scales with $\textrm{NA}^{-1}$ and the height scales with $\textrm{NA}^{-2}$. Hence an increase in $\textrm{NA}$ causes the bottle to become overall smaller and compressed along the $z$ direction.

%The waist and the Rayleigh range can be written in terms of the numerical aperture NA as
%\begin{eqnarray}
%\omega_0&=&\lambda_0/\pi \textrm{NA}\\
%z_R&=&n_{md}\lambda_0/\pi \textrm{NA}^2,
%\end{eqnarray}
%which mean that the width of the bottle scales with $\textrm{NA}^{-1}$ and the height scales with $\textrm{NA}^{-2}$. Therefore, the increase in $\textrm{NA}$ causes the bottle to become overall smaller, and to be compressed along the $z$ direction.

\subsection{Radiation forces}
When a Rayleigh particle is placed in an electromagnetic field, there are three forces that act on it \cite{Jones2015}. The first, called spin-curl force, is a result of polarisation gradients \cite{Albaladejo2009} and can be disregarded in the case of uniform linear polarization we are interested in. The second is called scattering force, and is proportional to the Poynting vector. Near the origin the scattering force points in the direction of propagation of the beam. Finally, the gradient force is proportional to the gradient of the potential energy of the particle under the influence of the electromagnetic field. From the intensity given by Eq. (\ref{eq:exact_intensity}), the scattering force $\vec{F}_{p}^{(scat)}(\vec{r})$, the gradient force $\vec{F}_{p}^{(grad)}(\vec{r})$ and the optical potential $V_p(\vec{r})$ acting on a trapped particle with radius $R$ and refractive index $n_p$ can be readily calculated using \cite{Li2013}
\begin{eqnarray}
 \vec{F}_{p}^{(scat)}(\vec{r}) &=& \hat{z}\frac{128\pi^5R^6}{3c\lambda_0^4}\left(\frac{m^2-1}{m^2+2}\right)^2n_{m}^5I_p(\vec{r})\\
 \vec{F}_{p}^{(grad)}(\vec{r})&=&\frac{2\pi n_{m}R^3}{c}\left( \frac{m^2-1}{m^2+2}\right)\nabla I_p(\vec{r})\\
\label{eq:potential_dipole}
 V_p(\vec{r})& =& -\frac{2\pi n_{m}R^3}{c}\left(\frac{m^2-1}{m^2+2} \right)I_p(\vec{r})
\end{eqnarray}
where $m=n_p/n_m$ is the particle-medium refractive index ratio. We are interested in situations in which the $ m $ parameter is smaller than $1$, in such a way that the particle is repelled by light.

\begin{figure}[t]
    \centering
    \includegraphics[width = \linewidth]{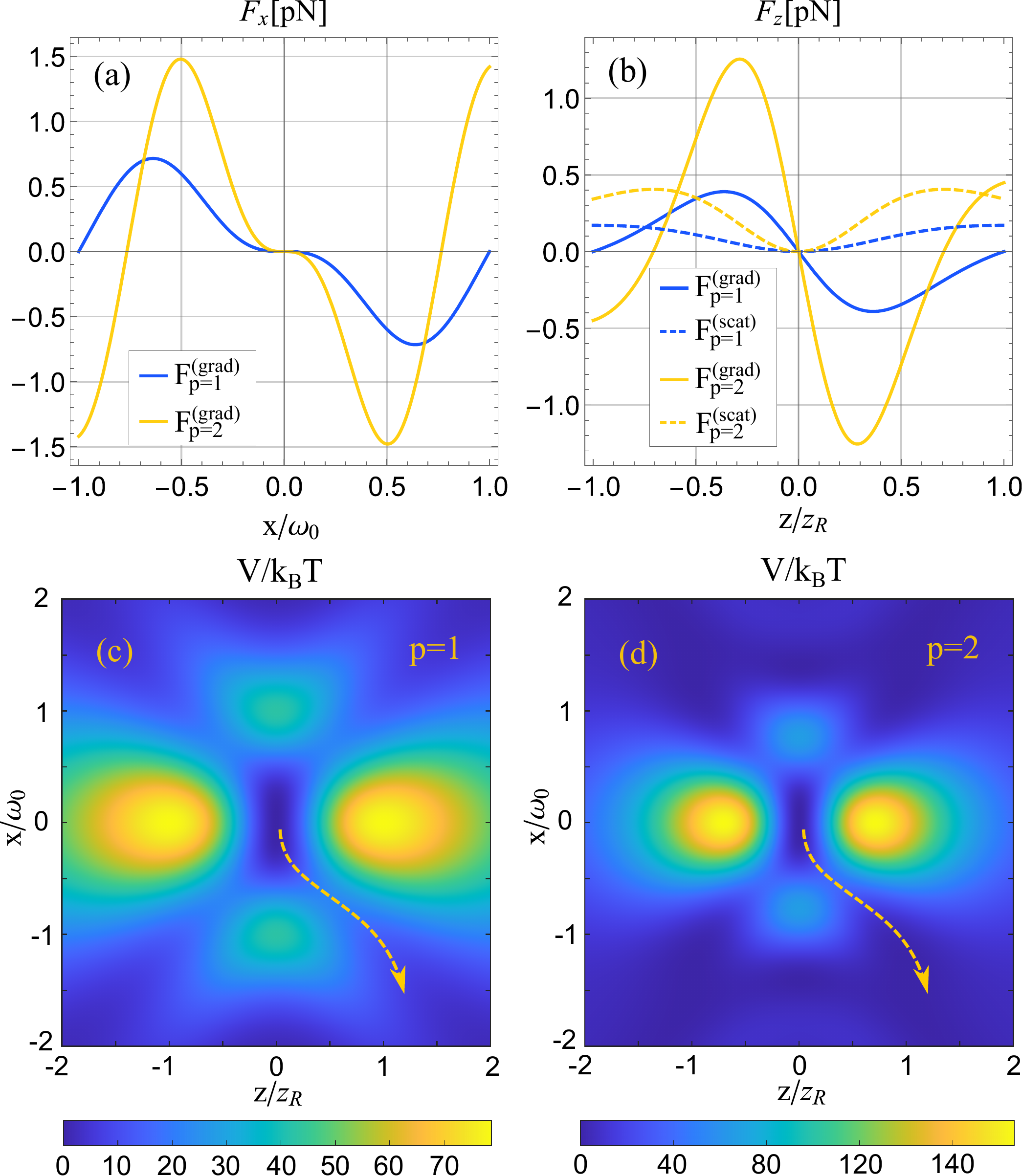}
    \caption{Forces acting on a trapped sphere in the (a) $x$ direction and (b) $z$ direction for $p=1$ and $p=2$. Solid lines are gradient forces, while dashed lines are scattering forces. Potential landscape in the $xz$ plane for a sphere trapped by a bottle beam with (c) $p=1$ and (d) $p=2$. The parameters used for these plots are: NA=0.5, $R=70$nm, $\lambda_0=780$nm, $n_m=1.46$, $n_p=1.33$, $P_0=200$mW, $T=300$K.}
    \label{fig:force_examples}
\end{figure}

The forces acting on a spherical water droplet ($n_p=$1.33) with \SI{70}{nm} radius trapped in oil ($n_m=$1.46) by a bottle beam ($\lambda=$\SI{780}{nm}, $P_0=$\SI{200}{mW} for each beam in the superposition) focused by an objective lens (NA=0.5) are displayed in Figure \ref{fig:force_examples}. As expected, the gradient forces point to the origin. Note that the scattering force, which points along the propagation direction, is null at the equilibrium position. This is in strong contrast to standard Gaussian traps and presents an advantage since the imbalance between scattering and gradient forces often poses challenges to optical trapping \cite{Nieminen2008}.

Another interesting feature of the bottle beam trap is the flat bottom of the intensity well in the $z=0$ plane, seen in Figure \ref{fig:intensity_examples}(a), and the approximate null derivative of the force along the radial direction at the origin. This can be understood by looking at the potential near the origin ($\rho\ll\omega_0$, $z\ll z_R$). It can be approximated to $4^{th}$ order as
\begin{equation}
    \frac{V_p(\rho,z)}{V_0}\approx \underbrace{\frac{4p^2}{\omega_0^4}\rho^4}_{T_{\rho^4}}-\underbrace{\frac{8p^2(p+1)}{\omega_0^2z_R^2}\rho^2z^2}_{T_{\rho^2 z^2}}+\underbrace{\frac{4p^2}{z_R^2}z^2}_{T_{z^2}},%-\frac{4p^2(p^2+5)}{3z_R^4}z^4
    \label{eq:intensity_approx}
\end{equation}
where $V_0=[2\pi n_mR^3(m^2-1)/c(m^2+2)]I_0$ and the term of order $\mathcal{O}((z /z_R)^4)$ has been neglected since $(z /z_R)^4\ll(z /z_R)^2$ for $z\ll z_R$. 
At the plane $z=0$ the potential scales with $\rho^4$. Therefore, the force scales with $\rho^3$ and has vanishing first and second derivatives. For $z\neq0$, Eq. \eqref{eq:intensity_approx} has a crossed term $\rho^2z^2$ that couples motion along the axial and radial directions. Because the scattering force is proportional to the intensity, it also has null derivatives at the equilibrium position and hence vanishes for a particle placed at and near the origin. 

Finally, the potential in the $xz$ plane is displayed in Figures \ref{fig:force_examples}(c) and \ref{fig:force_examples}(d) for the cases of $p=1$ and $p=2$. As it can be seen, a trapped particle does not need to go through the high intensity peaks along the $x$ or $z$ axis in order to escape the trap. Smaller potential barriers have to be climbed if the particle undergoes paths like the yellow dashed ones. We will call the lowest potential energy needed for the particle to leave the trap $V_{min}$. Because the potential scales with $V_0$, we have $V_{min}\propto V_0$.
%for which the highest potential energies are about 10\% ($p=1$) and 20\% ($p=2$) of the potential peak along the $z$ direction. 

\subsection{Decoupling approximation}
The axial and radial movements can be decoupled if the coupling term in Eq.(\ref{eq:intensity_approx}) is much smaller than the remaining terms. 
The conditions under which this assumption holds true can be found by estimating the magnitude of the particle's displacements under the influence of the trap. 
Neglecting the cross term and considering thermal equilibrium we may write
\begin{eqnarray}
\label{eq:rho4integral}
\langle \rho^4\rangle &=& \frac{1}{Z_0}\int d^3\vec{r} \rho^4 \exp{\left[-\frac{4V_0p^2}{k_BT}\left(\frac{\rho^4}{\omega_0^4}+\frac{z^2}{z_R^2}\right)\right]}\\
\label{eq:z2integral}
\langle z^2\rangle &=&\frac{1}{Z_0}\int d^3\vec{r} z^2 \exp{\left[-\frac{4V_0p^2}{k_BT}\left(\frac{\rho^4}{\omega_0^4}+\frac{z^2}{z_R^2}\right)\right]},
\end{eqnarray}
where $k_B$ is the Boltzmann constant, $T$ is the temperature and $Z_0$ is given by
\begin{equation}
\label{eq:Z0integral}
    Z_0 = \int d^3\vec{r} \exp{\left[-\frac{4V_0p^2}{k_BT}\left(\frac{\rho^4}{\omega_0^4}+\frac{z^2}{z_R^2}\right)\right]}.
\end{equation}

From Eqs. \eqref{eq:rho4integral}-\eqref{eq:Z0integral} we find that
\begin{eqnarray}
\label{eq:rho4}
    \sqrt[4]{\langle\rho^4}\rangle&=&\sqrt[4]{\frac{\omega_0^4k_BT}{8p^2V_0}}\\
    \label{eq:z2}\sqrt{\langle z^2\rangle}&=& \sqrt{\frac{z_R^2k_BT}{8p^2V_0}}.
\end{eqnarray}%$\sqrt[4]{\langle\rho^4}\rangle\approx\sqrt[4]{\omega_0^4k_BT/8p^2V_0}$ and $\sqrt{\langle z^2\rangle}\approx \sqrt{z_R^2k_BT/8p^2V_0}$. 

Although Eqs. (\ref{eq:rho4}) and (\ref{eq:z2}) were derived by neglecting the cross term, they can be used to estimate the magnitude of the three different terms in Eq. (\ref{eq:intensity_approx}). Through simple scaling we are led to
\begin{equation}
    \frac{T_{\rho^2z^2}}{T_{\rho^4}}\sim \frac{T_{\rho^2z^2}}{T_{z^2}}\sim  \frac{1+p}{\sqrt{2}p} \left(\frac{V_0}{k_BT}\right)^{-1/2}.
\end{equation} 
Because $V_{min}/k_BT\gg1$ is required for the particle to be confined in the presence of a thermal bath \cite{Li2013} and $V_{min}\propto V_0$, fulfillment of the decoupling condition is associated with increased trap stability.
%vc acha 

In the decoupling regime the optical potential becomes
%the ratio between the second and first terms - and between the third and second terms - scales approximately as $(p^2+1)/p\sqrt{V_0/k_BT}$. 
% is necessary for stable trapping this ratio is expected to be small for sufficiently small $p$.
\begin{equation}
    V_p(\rho,z)\approx \frac{k_\rho^{(3)}}{4}\rho^4+\frac{k_z}{2}z^2,
    \label{eq:intensity_approx2}
\end{equation}
with the constants $k_\rho^{(3)}$ and $k_z$ given by,
\begin{eqnarray}
\label{eq:k_rho}
k_\rho^{(3)}&=&\frac{64n_mP_0R^3}{c}\left(\frac{\pi{\rm{NA}}}{\lambda_0}\right)^6\left(\frac{1-m^2}{2+m^2}\right)p^2 \\
k_z&=&\frac{32P_0R^3\lambda_0^2}{\pi^2n_mc}\left(\frac{\pi{\rm{NA}}}{\lambda_0}\right)^6\left(\frac{1-m^2}{2+m^2}\right)p^2.
\end{eqnarray}

\subsection{Trapped particle dynamics}
\begin{figure}[t]
    \centering
    \includegraphics[width = \linewidth]{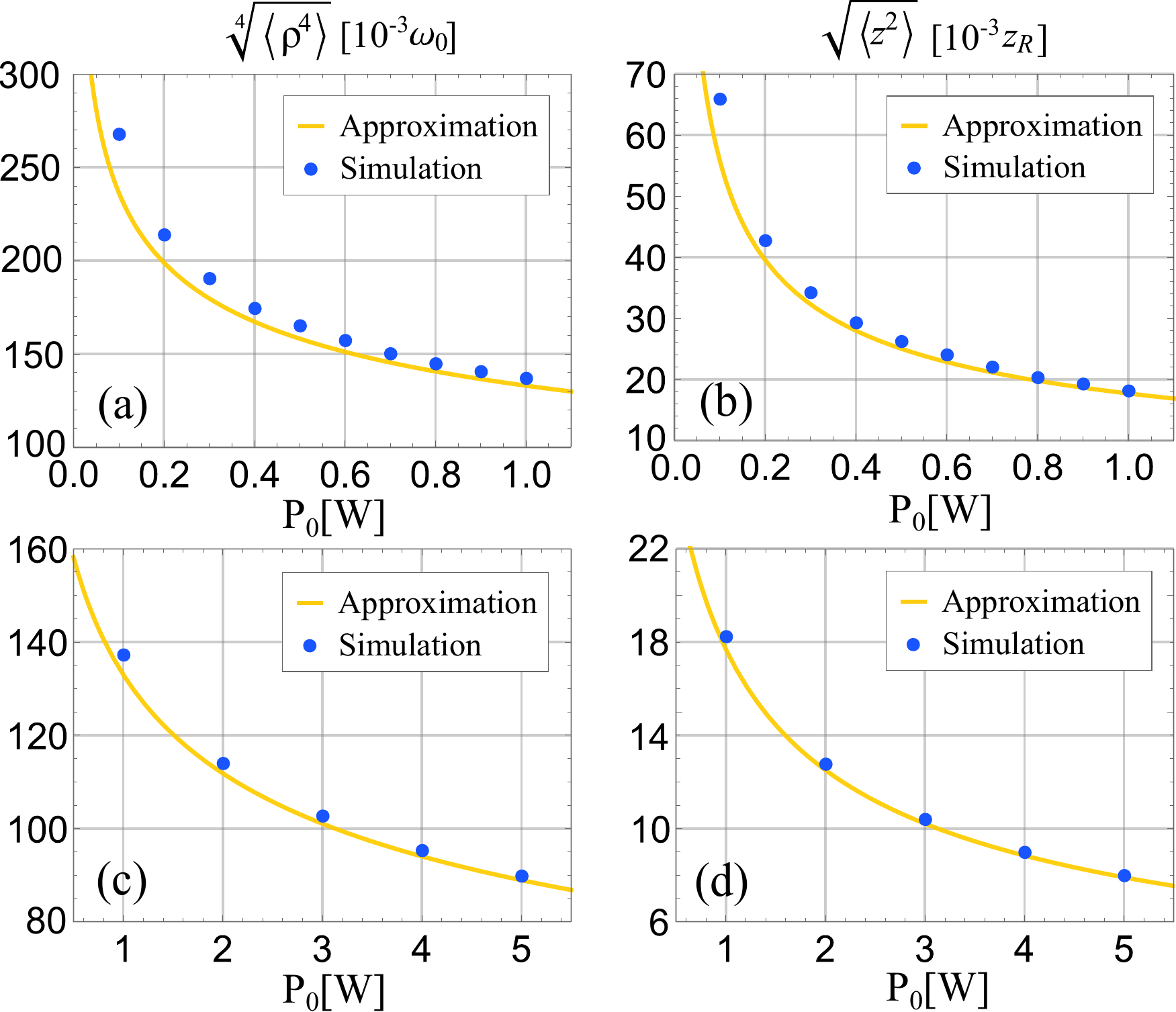}
    \caption{Comparison between the values of  $\sqrt[4]{\langle \rho^4\rangle}$ in (a),(c) and of $\sqrt{\langle z^2\rangle}$ in (b),(d) obtained from the approximated potential in Eq.(\ref{eq:intensity_approx2}) and from simulation of the particle's motion subject to the exact potential in Eq.(\ref{eq:potential_dipole}) for different laser powers. The motion was simulated during 10s with time steps of $0.5\mu$s using NA=0.5, $R=70$nm, $\lambda_0=780$nm, $n_m=1.46$, $n_p=1.33$, $T=295$K, $p=1$.
    }
    \label{fig:curves}
\end{figure}

To further evaluate the validity of the above estimates and approximations, it is useful to simulate the dynamics of a particle trapped by the potential of a bottle beam in its exact form, calculated from Eqs. (\ref{eq:exact_intensity}) and (\ref{eq:potential_dipole}). The equation of motion for a spherical particle under this condition is
\begin{equation}
    M\ddot{\vec{r}}(t)=-\gamma \dot{\vec{r}}(t)-\nabla V(\vec{r}(t))+\sqrt{2\gamma k_BT}\vec{W}(t),
    \label{eq:eom}
\end{equation}
where $\eta$ is the medium's viscosity,  $\gamma=6\pi\eta R$ is the drag coefficient and $M$ is the particle's mass. The environmental fluctuations are modelled using a Gaussian, white and isotropic stochastic process $\vec{W}(t)=(W_x(t), W_y(t), W_z(t))$, with zero mean and no correlations among different directions. We have that 
\begin{equation}
    \langle W_i(t) W_i(t') \rangle= \delta(t-t') \ ,
\end{equation}
where $\delta(t-t')$ is the Dirac delta in the time-domain.

For a sufficiently small particle the inertial term $M \ddot{\vec{r}}$ is negligible in comparison to the viscous term $\gamma \dot{\vec{r}}$. In this so-called \textit{over-damped} regime we can numerically integrate equation \eqref{eq:eom} using
%where $\gamma=6\pi\eta R$ is the drag coefficient, $m$ is the particle's mass and the last term is a stochastic force with $\vec{W}(t)=(W_x(t), W_y(t), W_z(t))$, $\langle W_i(t)\rangle=0$ and $\langle W_i(t)W_i(t')\rangle=\delta(t-t')$, $i=x,y,z$. Now, for $R$ sufficiently small, the inertial term in equation \ref{eq:eom} becomes negligible. In that case, the motion of the particle can be simulated iteratively using
\begin{equation}
    \vec{r}(t+\Delta t)=\vec{r}(t)-\frac{\nabla V(\vec{r}(t))}{\gamma}\Delta t+\sqrt{\frac{2k_BT\Delta t}{\gamma}}\vec{W}(t)
\end{equation}
where $\Delta t=\tau/n$ is the time interval between iterations, $\tau$ the total time of simulation and $n$ the total number of iterations. 

Numerical integration of the motion of a water droplet ($n_p=1.33$, $R=70$nm) trapped in oil ($n_m=1.46$) by a bottle beam ($p=1$, $\lambda=780$nm) focused using an objective lens (NA=0.5) was performed for different trapping powers. Note that the total trapping power is two times larger than the power $P_0$ of each beam. See Appendix A for details.

The motion was simulated for a period of 10s using time steps of 0.5$\mu$s. This resulted in $20\times10^6$ position values for each trapping power. The values of $\sqrt[4]{\langle x^4\rangle}$ and $\sqrt{\langle z^2\rangle}$ obtained from this simulation of the exact potential and the curves predicted using the approximated potential in Eq. \eqref{eq:intensity_approx2} together with Eqs. (\ref{eq:rho4integral}) and (\ref{eq:z2integral}) are displayed in Figure \ref{fig:curves}. The largest values of $\sqrt[4]{\langle x^4\rangle}/\omega_0$ and $\sqrt{\langle z^2\rangle}/z_R$ obtained are approximately 0.27 and 0.067, respectively. This justifies the fourth order approximation leading to Eq. \eqref{eq:intensity_approx} for the entire simulated range of trapping powers.

%The first thing we note is that for the simulated range of trapping powers, the largest values of $\sqrt[4]{\langle x^4\rangle}/\omega_0$ and $\sqrt{\langle z^2\rangle}/z_R$ obtained are, approximately, 0.27 and 0.067, which justifies the approximation to fourth order necessary for obtaining Eq.(\ref{eq:intensity_approx}).

%Next, we need to verify if neglecting the second term in Eq.(\ref{eq:intensity_approx}) and approximating the potential to that of Eq.(\ref{eq:intensity_approx2}) is possible. 

Moreover, we can see from Figure \ref{fig:curves} that agreement between the simulated dynamics of the exact potential and the approximate potential of Eq. \eqref{eq:intensity_approx2} increases with $P_0$.
%from Eq.(\ref{eq:intensity_approx2}) increases as the trapping power gets larger. 
For $P_0>1$W, exact and approximate values differ by less than 3\%, and hence Eq. \eqref{eq:intensity_approx2} can be considered a good approximation of the potential. 
This behavior is consistent with the previous estimate that the ratios $T_{\rho^2z^2}/T_{\rho^4}$ and  $T_{\rho^2z^2}/T_{z^2}$ scale with $V_0^{-1/2}$, and hence, the larger the trapping power the smaller the cross term in comparison to the remaining relevant terms. 

\begin{figure}[t]
    \centering
    \includegraphics[width = \linewidth]{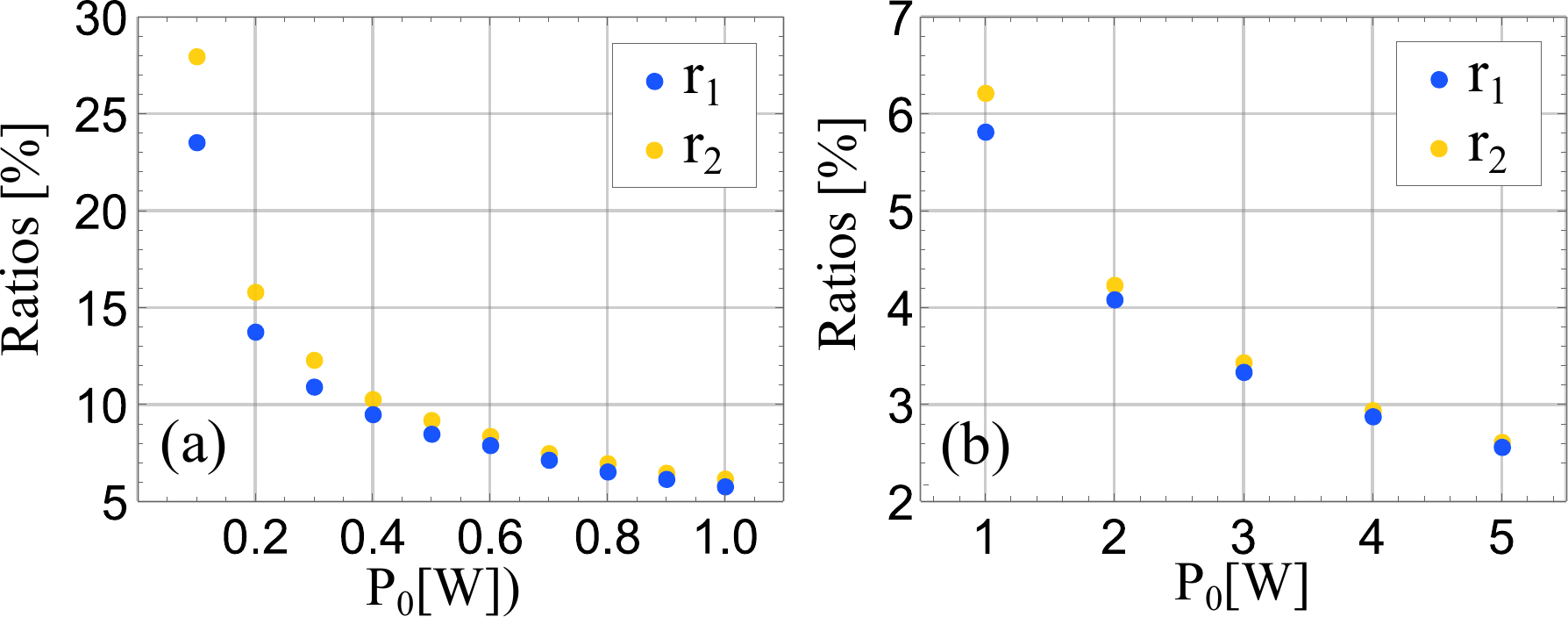}
    \caption{Ratios $r_1=\left\langle T_{\rho^2 z^2}\right\rangle/\left\langle T_{\rho^4}\right\rangle$ and $r_2 =  \left\langle T_{\rho^2 z^2}\right\rangle/\left\langle T_{z^2}\right\rangle$ obtained by simulating the motion of a particle subject to the exact potential in Eq.(\ref{eq:potential_dipole}) for different laser powers. The motion was simulated during 10s with time steps of $0.5\mu$s using NA=0.5, $R=70$nm, $\lambda_0=780$nm, $n_m=1.46$, $n_p=1.33$, $T=295$K.}
    \label{fig:ratios}
\end{figure}

This can be further verified in Figure \ref{fig:ratios}, where we plot the ratios 
%\begin{eqnarray}
%r_1 &=& \frac{\left\langle\frac{8p^2(p^2+1)}{\omega_0^2z_R^2}\rho^2z^2\right\rangle}{\left\langle\frac{4p^2}{\omega_0^4}\rho^4\right\rangle},\\
%r_2 &=& \frac{\left\langle\frac{8p^2(p^2+1)}{\omega_0^2z_R^2}\rho^2z^2\right\rangle}{\left\langle\frac{4p^2}{z_R^2}z^2\right\rangle},
%\end{eqnarray}
\begin{equation}
r_1 = \frac{\left\langle T_{\rho^2 z^2}\right\rangle}{\left\langle T_{\rho^4}\right\rangle}\,, \quad
r_2 =  \frac{\left\langle T_{\rho^2 z^2}\right\rangle}{\left\langle T_{z^2}\right\rangle},
\end{equation}
obtained from the simulations. The decreasing behavior of $r_1$ and $r_2$ with respect to $P_0$ confirms that increasing the trapping power is an effective way of decoupling the radial and axial directions. 

\begin{figure}[t]
    \centering
    \includegraphics[width =  \linewidth]{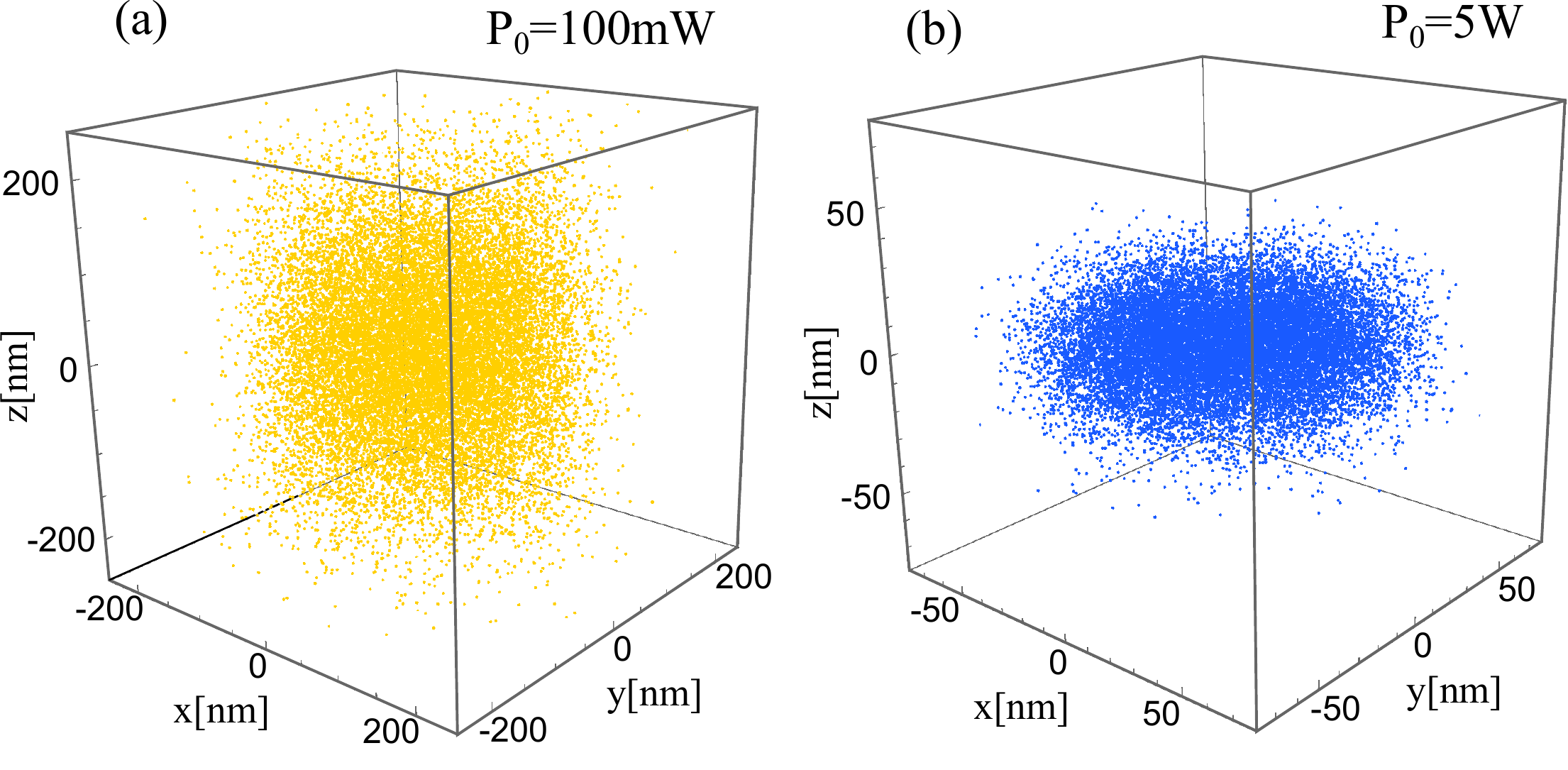}
    \caption{Positions of the particle obtained by simulating the motion of a particle trapped by the exact potential in Eq.(\ref{eq:potential_dipole}) for (a) $P_0=100$mW and (b) $P_0=5$W. The motion was simulated during 10s with time steps of $0.5\mu$s using NA=0.5, $R=70$nm, $\lambda_0=780$nm, $n_m=1.46$, $n_p=1.33$, $T=295$K, $p=1$. To allow better visualization, the $20\times10^6$ positions generated by the simulation were divided in 1000 sets, and only the first value of each set is displayed in the figure.}
    \label{fig:motion}
\end{figure}

Another consequence of the interplay between a radial quartic and a longitudinal quadratic potential is that elongation of the trap can be adjusted by tuning the laser power. This is illustrated in Figure \ref{fig:motion}, in which the positions of the trapped particle obtained from the numerical simulation are displayed in a scatter plot, for $P_0=100$mW and $P_0=5$W. As it can be seen, the trap is appreciably compressed along the z axis in the latter case, but not in the former. This feature is not present in regular Gaussian tweezers: since the potential is quadratic along the three axis, the expected value of the displacement along all axes scale equally with $\sqrt{P}$. %implying a trap elongation independent from trapping power. 
%Another consequence of a radial quartic term in the potential in the radial plane while having a quadratic one in the longitudinal axis is that the elongation of the trap can be adjusted by adjusting the trapping power. This illustrated in Figure \ref{fig:motion}, in which the simulated positions of the trapped particle are displayed for trapping powers of 100mW and 5W. As it can be seen, the trap is compressed along the z axis in the latter case, but not in the former. This feature is not present in regular Gaussian traps: since the potential is quadratic along the three axis, the expected value of the displacement along each axis scales with $\sqrt{P}$, implying a trap elongation independent from trapping power. 

We note that this compression is different from the one caused by an increase in numerical aperture, mentioned previously. In that case we have a compression of the overall shape of the intensity landscape along the $z$ axis, which happens in the case of a Gaussian beam due to the scaling of $\omega_0$ with $\textrm{NA}^{-1}$ and of $z_R$ with $\textrm{NA}^{-2}$. In contrast, an increase in the trapping power of a bottle beam compresses the region visited by the particle over time.

%, we discussed the compression of the overall shape of the intensity landscape along the $z$ axis, which also happens in the case of a Gaussian beam due to the scaling of $\omega_0$ with $\textrm{NA}^{-1}$ and of $z_R$ with $\textrm{NA}^{-2}$. Here, the increase of trapping power compresses the region visited by the particle during its motion.

%The effect of the null values of the force in the radial direction and of its two first derivatives can be seen clearly in figure \ref{fig:motion}, in which the points obtained in the simulation for 5W of trapping power are displayed: the particle is much more confined in the axial direction when compared to the radial direction. This behaviour differs from regular Gaussian traps, which are typically elongated along the $z$ axis.

%\begin{figure}[t]
%    \centering
%    \includegraphics[width =  \linewidth]{Robust_motion_alt.pdf}
%    \caption{Positions of the particle obtained by simulating of the motion of a particle trapped by the exact potential in Eq.(\ref{eq:potential_dipole}): (a) in the $xy$ plane for $P_0=100$mW; (b) in the $xz$ plane for $P_0=100$mW; (c) in the $xy$ plane for $P_0=5$W; (d) in the $xz$ plane for $P_0=5$W. The motion was simulated during 10s with time steps of $0.5\mu$s using NA=0.5, $R=70$nm, $\lambda_0=780$nm, $n_m=1.46$, $n_p=1.33$, $T=295$K, $p=1$. To allow better visualization, the $2\times10^6$ positions generated by the simulation were divided in 1000 sets, and only the first value of each set is displayed in the figure.}
 %   \label{fig:motion}
%\end{figure}

%A necessary condition for the quantities $z$ and $\rho$ to be independent is th
%at they have null correlation, i.e.
%\begin{equation}
%    \mathbb{C}(\rho,z) = \frac{\langle (\rho-\langle \rho\rangle)(z-\langle z\rangle)\rangle}{\sqrt{\langle \rho^2\rangle - \langle\rho\rangle^2}\sqrt{\langle z^2\rangle-\langle z\rangle^2}}=0.
%\end{equation}

In summary, we conclude that in the dipole regime Eq. \eqref{eq:intensity_approx} is a good approximation for the optical potential generated by a bottle beam for a wide range of trapping powers, as it only relies on $\rho^4/\omega_0^4\ll1$ and $z^4/z_R^4\ll1$. 
On the other hand, decoupling of the radial and axial motions only occurs for high trapping powers that make the cross term in Eq. \eqref{eq:intensity_approx} negligible. This allows approximating the potential by Eq. \eqref{eq:intensity_approx2}. Furthermore, increasing the trapping power causes squashing along the axial direction of the accessible region for a trapped particle.

%\iffalse

\subsection{Decoupling by addition of an extra mode}
An alternative way to decouple the radial and longitudinal dependencies of the potential is to add an extra Laguerre-Gauss mode with $\ell_2=0$ and $p_2\neq0$ to the superposition. Consider the intensity $I(\rho, z)$ of the following superposition: 
\begin{eqnarray}
\label{}
\hspace{-4mm}E(\rho, z)\!=\!E^{LG}_{0, 0}(\rho,z)\!+\! \alpha_1E^{LG}_{0,1}(\rho,\!z)\!+\!\alpha_2 E^{LG}_{0,p_2}(\rho,z),
\end{eqnarray}
where $\alpha_{j=1,2}$ are complex amplitudes. The condition to have a bottle beam is that $I(\rho, z)$ vanishes at the focus. The light intensity at the beam focus is proportional to
\begin{eqnarray}
\label{eq:bottle_condition}
I(\mathbf{0}) = \mathcal{N}^2 
\big|1 + \alpha_1 + \alpha_2\big|^2,
\end{eqnarray}
where $\mathcal{N}=\sqrt{4P_0/c\epsilon\pi\omega_0^2}$ .

With the appropriate approximations (see Appendix B for details) and the bottle beam condition in Eq.\eqref{eq:bottle_condition}, we obtain the approximate intensity $I(\rho, z)$,
\begin{eqnarray}
\!\!\!\!\mathcal{N}^2\!\!\left[
4|B|^2\!\left(\frac{z^2}{z_R^2} \!+\!\frac{\rho^4}{w_0^4}\right)\!\!-8\!\left[|B|^2\!+\!\!\mathrm{Re}\left(AB^*\right)\right]\!\frac{z^2\!\rho^2}{z_R^2w_0^2}\right]\!\!\!\,,
\end{eqnarray}
where $A = \alpha_1 + \alpha_2 p_2^2$ and $B = \alpha_1 + \alpha_2p_2$.
Decoupling of the radial %$(\rho)$ 
and longitudinal %$(z)$
motions can then be achieved by choosing $\alpha_1$ and $\alpha_2$
such that
\begin{eqnarray}
|B|^2 + \mathrm{Re}\left(AB^*\right)=0\qquad (B\neq 0).
\end{eqnarray}

As an example, let us choose $p_2=2$. A decoupled bottle beam can be obtained by the superposition coefficients
\begin{eqnarray}
\alpha_1 &=& -3/2, \\  \alpha_2 &=& 1/2.
\end{eqnarray}
Using Eq.\eqref{eq:potential_dipole} we can find the decoupled potential near the origin ($\rho\ll\omega_0$, $z\ll z_R$). This is given to $4^{th}$ order by
\begin{eqnarray}
\label{eq:potencial_dipole_2}
V(\rho, z) \approx V_0
\left(\frac{z^2}{z_R^2} + \frac{\rho^4}{w_0^4}\right),
\end{eqnarray}
%where $V_0=[2\pi n_mR^3(m^2-1)/c(m^2+2)]I_0$.
which has the same form as Eq.(\ref{eq:intensity_approx2}). Moreover, this solution yields the maximum trap stiffness of the three mode configuration, as demonstrated in Appendix B. 

\subsection{Calibration of the optical trap}
In laboratory conditions, quantitative measurements using optical tweezers rely on knowledge of the trap's parameters. In the case of a bottle trap defined by the potential in Eq. \eqref{eq:intensity_approx2}, the relevant parameters are $k_z$ and $k^{(3)}_\rho$. 
To properly operate the tweezer these must be found by measuring the particle's position during a finite interval of time. This yields a time series $\vec{r}_m(t)=\vec{\beta}\cdot\vec{r}(t)$, where $\vec{\beta}=(\beta_x, \beta_y, \beta_z)$ are conversion factors between position displacements and the measured quantity, such as the voltage in a position sensitive detector. 
For simplicity, we will assume $\beta_x=\beta_y=\beta_\rho$.
%, as this is the case for most measurement schemes \cite{alguma_citacao_aqui}. 
%Full calibration requires knowledge of these factors as well.
%Acho que essa frase aqui em cima nao e necessaria.

For a bottle beam trap the particle's position can be measured using a high speed camera \cite{Gibson2008}, or alternatively by applying a purely Gaussian beam at a different wavelength with respect to the bottle beam. The second beam can be focused onto the trapped particle by the same objective lens used for the bottle, and collected by a second objective lens after separation from the trapping beam by a dichroic mirror. The collected Gaussian light can then be directed onto a Quadrant Photo Detector, where the usual forward scattering measurement is performed \cite{Pralle1999}. The Gaussian power should be kept significantly weaker then the Bottle power to avoid disturbances due to the presence of this auxiliary Gaussian trap.

In the decoupled regime, movement along the $z$ axis is independent from movement along the $x$ and $y$ axes and the equations of motion can be separated from Eq. \eqref{eq:eom}, yielding
\begin{equation}
    -\gamma \dot{z}(t)-k_zz(t)+\sqrt{2\gamma k_BT}W_z(t)=0,
\end{equation}
where once again we assume the inertial term is negligible. The constants $k_z$ and $\beta_z$ can be found using the standard procedure of analysing the autocorrelation function \cite{Alves2012} or the power spectral density \cite{BergSorensen2004, Melo2020} of the measured axial displacements $z_m(t)=\beta_zz(t)$. 

To find the remaining relevant constants we need two independent equations. Using Eqs.(\ref{eq:rho4integral}) and (\ref{eq:k_rho}) we may write
\begin{equation}
    \frac{k_\rho^{(3)}\langle\rho^4\rangle}{4} =  \frac{k_BT}{2}\rightarrow\langle\rho^4\rangle = \frac{2k_BT}{k_\rho^{(3)}},
\end{equation}
leading to the relation,
\begin{equation}
\label{eq:calibrateX1}
    \langle\rho_m^4\rangle = \beta_\rho^4 \frac{2k_BT}{k_\rho^{(3)}}.
\end{equation}

A second equation can be obtained from an active method of calibration consisting of moving the sample in which the particle is immersed with a known velocity $\vec{v}_{drag}$ \cite{Simmons1996, Brouhard2003}. This will cause a constant drag force $\gamma \vec{v}_{drag}$ on the particle, and taking $\vec{v}_{drag}=v_{drag}\hat{x}$ the equation of motion along the $x$ axis becomes
\begin{equation}
\label{eq:eomX}
    \gamma v_{drag}-\gamma \dot{x}(t)-k_\rho^{(3)}x(t)\rho(t)^2+W_x(t)=0 \ .
\end{equation}
After a transient time the particle reaches an equilibrium position displaced with respect to the trap's center, with $\langle \dot{x}(t)\rangle = 0 $. Taking the time average of Eq.(\ref{eq:eomX}) leads to
\begin{equation}
    \gamma v_{drag}-k^{(3)}_\rho\langle x(t)\rho(t)^2\rangle=0,
\end{equation}
which can then be used to obtain the relation
\begin{equation}
\label{eq:calibrateX2}
\langle x_m(t)\rho_m(t)^2\rangle = \beta_\rho^3 \frac{\gamma v_{drag}}{k_\rho^{(3)}}.
\end{equation}

Eqs.(\ref{eq:calibrateX1}) and (\ref{eq:calibrateX2}) together with the standard autocorrelation procedure for the axial motion enables the measurement of the four parameters $\beta_\rho,\beta_z, k_\rho^{(3)}$ and $k_z$ in the decoupled approximation.

\section{Intermediate regime}
In many applications it is desirable to trap `large' micron-sized particles such as living cells \cite{Zhong2013, Liang2019}. 
This presents an intermediate regime, in which the size of the particle is comparable to the wavelength of the trapping beam ($R\approx\lambda$) and neither the dipole ($R\ll\lambda$) nor geometric optics ($R\gg\lambda$) approximations can be used to calculate the optical forces. 
Instead, the forces must be calculated using the so-called generalized Lorenz–Mie theory (GLMT), for which we provide a brief introduction following the treatment presented in \cite{Nieminen2007}.

%In this intermediate regime, the size of the particle is comparable to the wavelength of the trapping beam ($R\approx\lambda$) and neither the dipole approximation ($R\ll\lambda$) nor geometric optics ($R\gg\lambda$) can be used to calculate the optical forces. 
%Instead, the forces must be calculated using the generalized Lorenz–Mie theory (GLMT). These calculation can be rather complex, and here we only introduce them briefly following the treatment presented in \cite{Nieminen2007}.

\begin{figure*}[t]
    \centering % \textwidth % \linewidth
    \includegraphics[width=\linewidth]{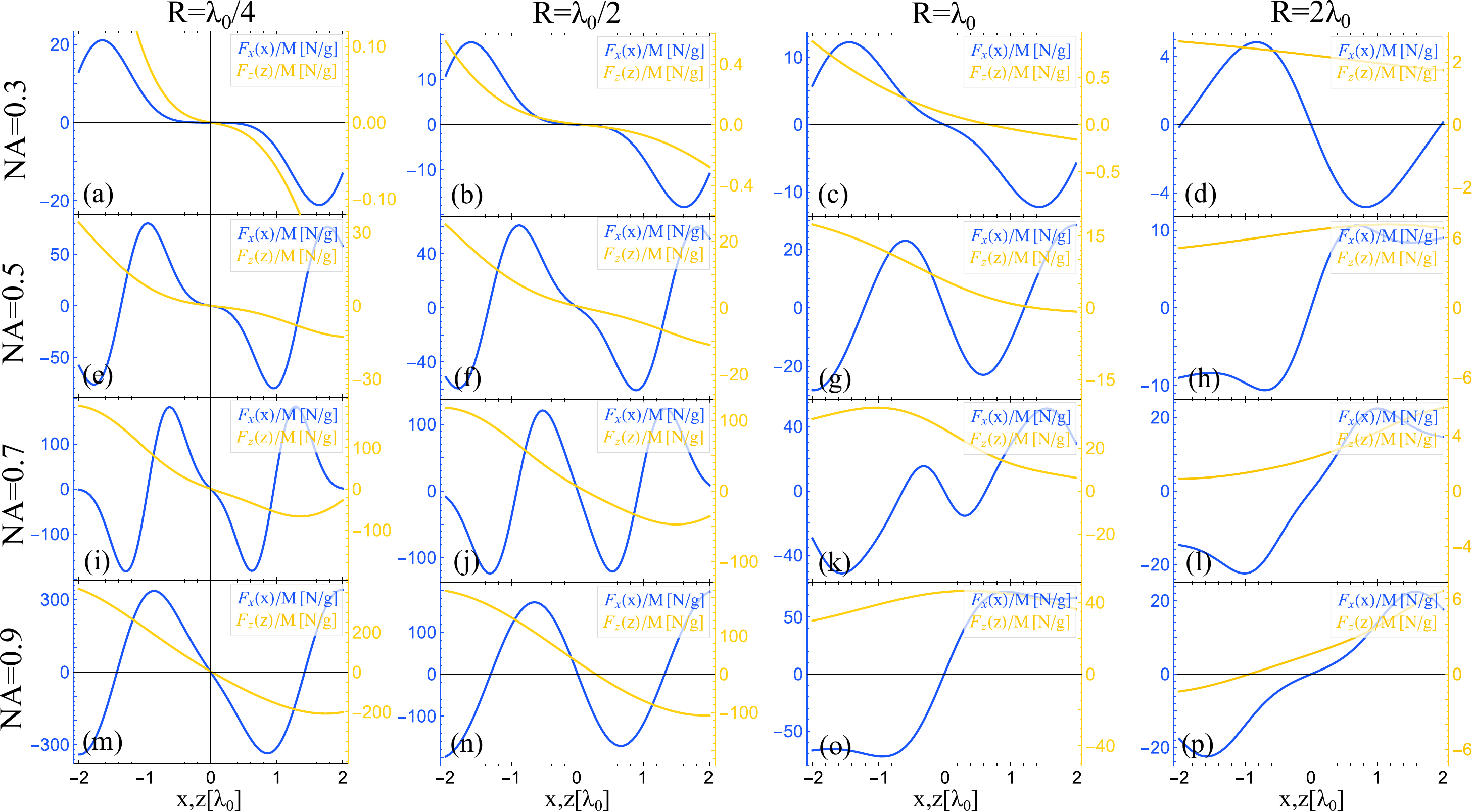}
    \caption{Optical forces acting on a particle trapped by a bottle beam in the intermediate regime ($R\approx\lambda$). The force in the axial direction ($F_z(z)$) is calculated for $x=y=0$, while the force in the radial direction ($F_x(x)$) is calculated for $y=0$ and $z=z_{eq}$. If no axial equilibrium position is found, $F_x(x)$ is evaluated at $z=0$. The particle's radius is constant within each column, while the numerical aperture is constant within each line. The forces are normalized by the particle's mass, other parameters used in simulation are: $n_p=1.33$, $n_m=1.46$, $\lambda_0=780$nm, $P=500$mW, density of the particle$=10^3kg/m^3$, $p=1$.}
    \label{fig:many_intermediate}
\end{figure*}

\subsection{Generalized Lorenz-Mie Theory}
Regardless of the size of the trapped particle, optical forces arise from the exchange of momentum with the photons from the trapping beam. Therefore, the total momentum transferred to the particle is equal to the change in momentum of the scattered electromagnetic field. It is then useful to separate the field in incoming $\vec{E}_{in}$ and outgoing $\vec{E}_{out}$ parts, which in turn can be expanded in terms of vector spherical wave-functions (VSWFs) defined in a coordinate system centered at the particle's center,
\begin{eqnarray}
\label{eq:expasion_ein}
\vec{E}_{in} = \sum^\infty_{i=1}\sum^i_{j=-i}a_{ij}\vec{M}_{ij}^{(2)}(k\vec{r})+b_{ij}\vec{N}^{(2)}_{ij}(k\vec{r}),\\
\label{eq:expasion_eout}
\vec{E}_{out} = \sum^\infty_{i=1}\sum^i_{j=-i}p_{ij}\vec{M}_{ij}^{(1)}(k\vec{r})+q_{ij}\vec{N}^{(1)}_{ij}(k\vec{r}),
\end{eqnarray}
where $\vec{M}_{ij}^{(1)},\vec{N}_{ij}^{(1)},\vec{M}_{ij}^{(2)}$ and $\vec{N}_{ij}^{(2)}$ are the VSWFs, with the upper index (1) standing for outward-propagating transverse electric and transverse magnetic multipole fields and (2) for the corresponding inward-propagating multipole fields.

The coefficients $a_{ij}$ and $b_{ij}$ can be calculated for the incident beam and used to obtain the $p_{ij}$ and $q_{ij}$ coefficients for the scattered field by a simple matrix-vector multiplication between the so-called $T$-matrix and a vector containing the coefficients of the incoming field. The $T$-matrix depends only on the characteristics of the trapped particle, which we assume spherical. Once the coefficients are calculated, the force along the axial direction $z$ is given by
\begin{eqnarray}
\label{eq:force_z_mie}
    F_z&=&\frac{2n_{md}P}{cS}\sum_{i=1}^\infty\sum_{j=-1}^i\frac{j}{i(i+1)}\textrm{Re}(a_{ij}^*b_{ij}-p_{ij}^*q_{ij})-\nonumber\\&&\frac{1}{i+1}\sqrt{\frac{i(i+2)(i-j+1)(i+j+1)}{(2i+1)(2i+3)}}\times\nonumber\\
    &&\hspace{-1mm}\textrm{Re}(a_{ij}a_{i+1,j}^*\hspace{-1mm}+\hspace{-1mm}b_{ij}b_{i+1,j}^*\hspace{-1mm}-\hspace{-1mm}p_{ij}p_{i+1,j}^*\hspace{-1mm}-\hspace{-1mm}q_{ij}q_{i+1,j}^*)
\end{eqnarray}
with
\begin{equation}
    S=\sum_{i=1}^\infty\sum_{j=-i}^i(\vert a_{ij}\vert^2+\vert b_{ij}\vert^2).
\end{equation}

Forces acting along the $x$ and $y$ axis have more complicated formulae and can be more easily calculated by rotating the coordinate system. The effect of displacing the particle can be taken into account by appropriate translations of the trapping beam. 

Due to the linearity of Eqs. \eqref{eq:expasion_ein} and \eqref{eq:expasion_eout}, the expansion coefficients for a superposition of different beams can be found by adding the expansion coefficients for each beam, and subsequently substituted in Eq. \eqref{eq:force_z_mie} to calculate the resultant force. 
We shall use the latest version of the toolbox developed in \cite{Nieminen2007} to perform these computations for the case of a particle trapped by a bottle beam.

\subsection{Optical forces from a bottle beam}

Optical forces generated by the superposition of a Gaussian beam and a Laguerre-Gauss beam with $\ell=0, p\neq0$ are obtained with the aid of \cite{Nieminen2007, Lenton2020}.
For simplicity, we focus on the $p=1$ case and a particle of refractive index $n_{p}=1.33$ trapped by a 500 mW beam at $\lambda_0=780$ nm immersed in oil of refractive index $n_{m}=1.46$.

Figure \ref{fig:many_intermediate} shows the plots of $F_z(z)$ and $F_x(x)$ divided by the particle's mass for four different NA's and four different particle radii. The force in the $z$ direction is evaluated for $x=y=0$, while $F_x(x)$ is evaluated at $y=0, z=z_{eq}$, where $z_{eq}$ is the equilibrium coordinate along the $z$ direction, i.e.,
\begin{equation}
    \begin{cases}
    F_z(z_{eq})=0\\
    dF_z(z)/dz\vert_{z=z_{eq}}<0
    \end{cases}
\end{equation}
When no equilibrium position exists, $F_x(x)$ is evaluated at $z=0$.  

Some general trends can be extracted from Figure \ref{fig:many_intermediate}. First, we note that if the sphere is small ($R=\lambda_0/4$) and the numerical aperture is low (NA $=0.3,0.5$), the force in the $x$ direction resembles the one calculated using the dipole approximation, i.e., it appears to scale with $x^3$ around the origin. As R or NA increases, this cubic dependence starts to vanish, giving place to a linear dependence.

%For NA = 0.5, as the particle gets larger, this cubic dependence starts to vanish, giving place to a linear dependence for $R=\lambda_0$ and finally turning into a non-restorative force for $R=2\lambda_0$.

We can also notice that the size of the particle and the numerical aperture play an important role on the existence of an equilibrium position in the axial direction, with large radius $R$ and large NA being detrimental to the trap stability along the $z$ axis. For NA = 0.7, for instance, there is an equilibrium position if $R=\lambda_0/4$ or $R=\lambda_0/2$, but not if $R$ is larger. For a fixed $R=\lambda_0$, $z_{eq}$ doesn't exist for NA $>0.5$. This is rather different from what happens in the regular Gaussian trap, in which increasing the NA is associated with an increase in trap stability \cite{Li2013}. 

\subsection{Limitations of trapping in a dark focus}
The trends observed in Figure \ref{fig:many_intermediate} can be understood qualitatively by recalling that a bottle beam is a dark region surrounded by a finite bright light boundary. If the particle is small enough it will fit inside the dark region and will be repelled by the boundary. 
In contrast, if the particle is too big it does not fit inside the bottle and the dark focus becomes irrelevant, with the beam effectively pushing the particle away.

This can be seen for in Figures \ref{fig:many_intermediate}(e)-(h): the dimensions of the bottle when NA $=0.5$ are $W=0.99\mu$m and $H=2.9\mu$m. Therefore, a particle of diameter $0.5\lambda$ fits entirely inside the bottle and is free within the dark region, causing the force in the $x$ direction to have vanishing derivative near the origin. When $R=\lambda_0$, the particle no longer fits in the dark focus, and the influence of light gives a linear scaling to $F_x(x)$ around the equilibrium position. When $R=2\lambda_0$ the particle has an increased overlap with the light intensity and no longer encounter an equilibrium position.

\begin{figure}[t]
    \centering
    \includegraphics[width = \linewidth]{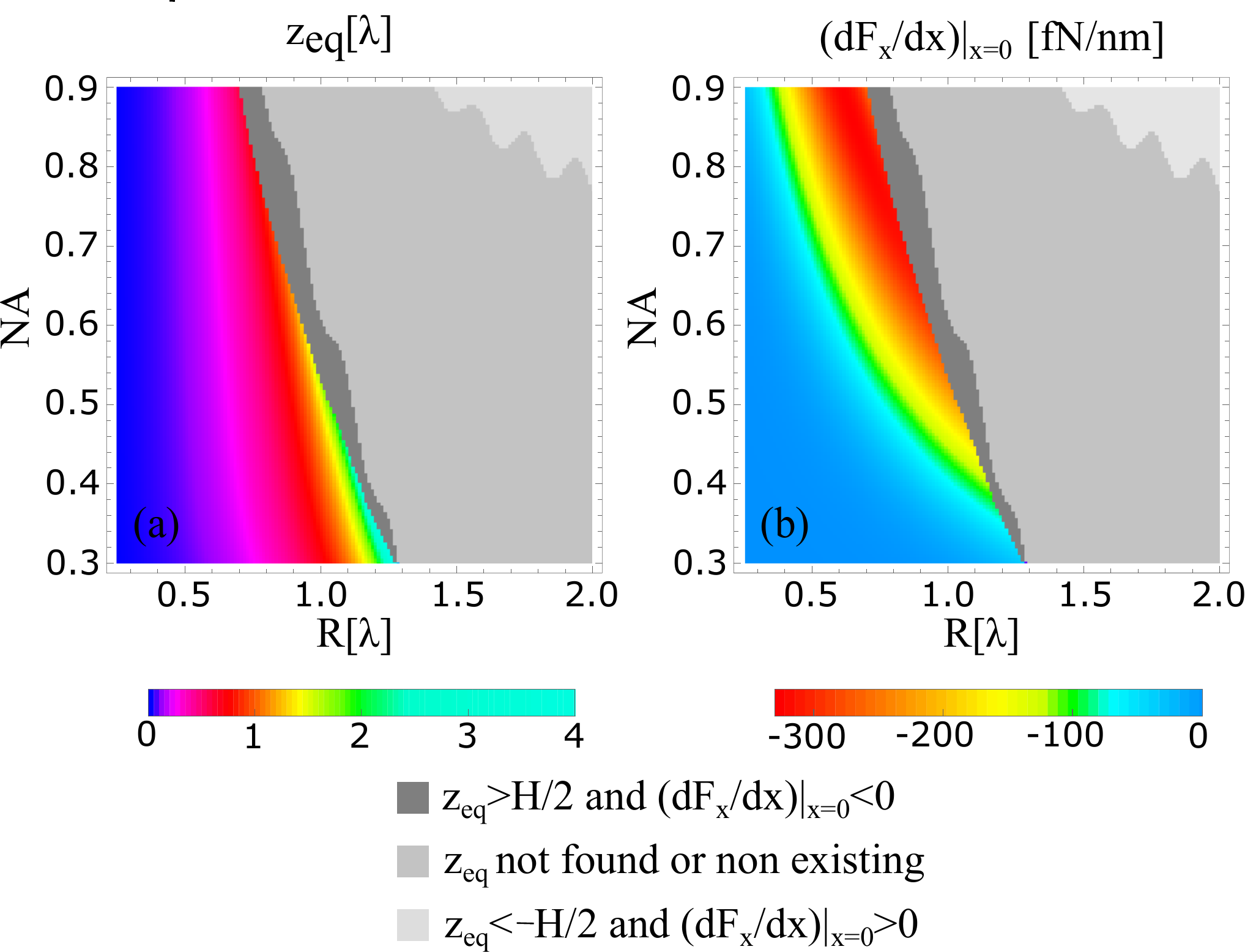}
    \caption{Intermediate regime simulations for different values of NA and $R$: (a) Axial equilibrium coordinate and (b) first derivative of the force in the radial direction. Medium gray: no equilibrium position was found in the inspected range ($-6\lambda_0<z<6\lambda_0$). Light (dark) gray: an equilibrium position was found outside the bottle, at $z_{eq}<-H/2$ ($z_{eq}>H/2$) and the force is non-restorative (restorative) along the radial direction. The regions in different colors are the ones in which trapping inside the bottle is possible. The parameters used in the simulation were $\lambda_0=780$nm, $P=500$mW, $n_m=1.46$, $n_p=1.33$, $p=1$.}
    \label{fig:limiting}
\end{figure}

Similarly, an increase in numerical aperture causes the dark focus to shrink. When the bottle becomes too small to comprise the particle the situation in the third column of Figure \ref{fig:many_intermediate} is reached and the forces eventually turns into non-restorative ones.

This qualitative reasoning is confirmed in Figure \ref{fig:limiting}, in which the equilibrium position $z_{eq}$ and the derivative along the $x$ direction near the equilibrium position are displayed as a function of the particle's radius and the numerical aperture. Two main regions can be identified in each of the plots. The first of them, is the region for which $z_{eq}$ was not found in the range of inspected axial coordinates $-6\lambda_0<z<6\lambda_0$. In this region, the derivative along the $x$ axis was not evaluated. The remaining areas are the ones in which an axial equilibrium position exists.

Because we wish to trap the particle in the dark focus, we need to avoid equilibrium situations as the ones described in \cite{Gahagan1996} in the context of vortex beams, in which the scattering force is balanced by the repelling gradient force before the focus. To exclude trapping positions outside the bottle, the regions in which $z_{eq}>H/2$ were displayed in dark grey and the regions in which $z_{eq}<-H/2$ were displayed in light grey. In the latter case, the derivative of the radial force was found to be positive and hence non-restorative, while in the former this derivative was found to be negative. The coloured region, then, is the region for which stable trapping inside the bottle is possible.

%The axial equilibrium positions displayed in Figure \ref{fig:limiting} can reach values near $6\lambda_0$. In order to trap particles inside the bottle beam, we must demand that $z_{eq}$ lies in the interval $(-H/2, H/2)$. Since we are considering the case $p=1$, this condition is equivalent to $\vert z_{eq}\vert<z_R$. Figure \ref{fig:limiting} shows the region in which trapping inside the bottle is possible, considering also the need for a restorative force in the radial direction.

\begin{figure}[t]
    \centering
    \includegraphics[width=\linewidth]{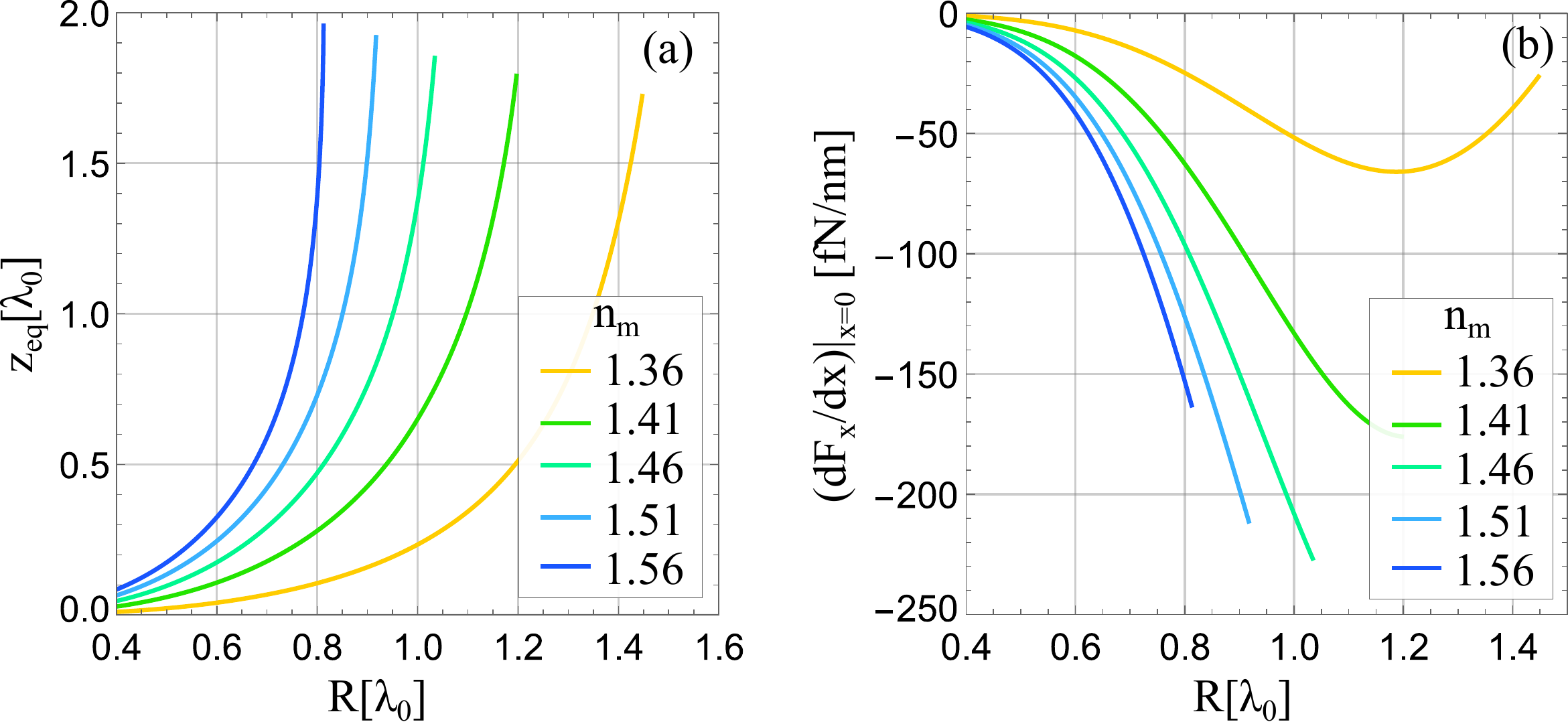}
    \caption{Intermediate regime simulations for different values $n_m$: (a) Axial equilibrium coordinate and (b) first derivative of the force in the radial direction as a function of particle's radius. Points for which $z_{eq}>H/2$ are not displayed. The parameters used in the simulation were $\lambda_0=780$nm, $P=500$mW, $n_m=1.46$, $n_p=1.33$, $p=1$.}
    \label{fig:refractive}
\end{figure}

We can then conclude that for a given $R$, there is a maximum numerical aperture that can be used to form a stable trap. Conversely, for a given NA, there is a limit on the size of the particles that can be trapped. Figure \ref{fig:refractive}(a) shows how this limit varies for different refractive indices of the medium and a fixed NA $=0.5$. The curves were chopped when $z_{eq}$ became larger than $H/2$, and we can clearly see that the closer the refractive index gets to that of the particle, the larger the radius of the particle that can be trapped. Figure \ref{fig:refractive} confirms that the radial force is restorative for the entire range of $R$ and $n_m$ we considered. It also shows that while decreasing $n_m$ can help trapping larger particles, it also diminishes the force experienced by the particle, and hence plays an important role in the trap's stability. 

\subsection{Trapping living organisms in the dark}

The bottle beam trap finds promising applications in biology. For instance it has been reported that organelles with a refractive index lower than its surroundings are repelled from standard Gaussian optical tweezers \cite{Zhang2008}. The bottle beam could then be used to manipulate such organelles within a cell. 

Similarly, a dark optical trap could also be employed to trap living organisms without excessive laser damage onto the cell by appropriate choice of a surrounding medium. Iodixanol has been reported as a non-toxic medium for different organisms, with high water solubility, in which the refractive index can be linearly tuned in the visible to near-IR range from $\sim1.33$  to $\sim 1.40$ by changing concentration \cite{Boothe2017}. Assuming a mean refractive index for a living cell to be within the range  $\sim1.36$  to $\sim 1.39 $ \cite{Liu2016, Rappaz2005} it is expected that stable trapping in a dark focus can be attained.

\begin{figure}[t]
    \centering
    \includegraphics[width=\linewidth]{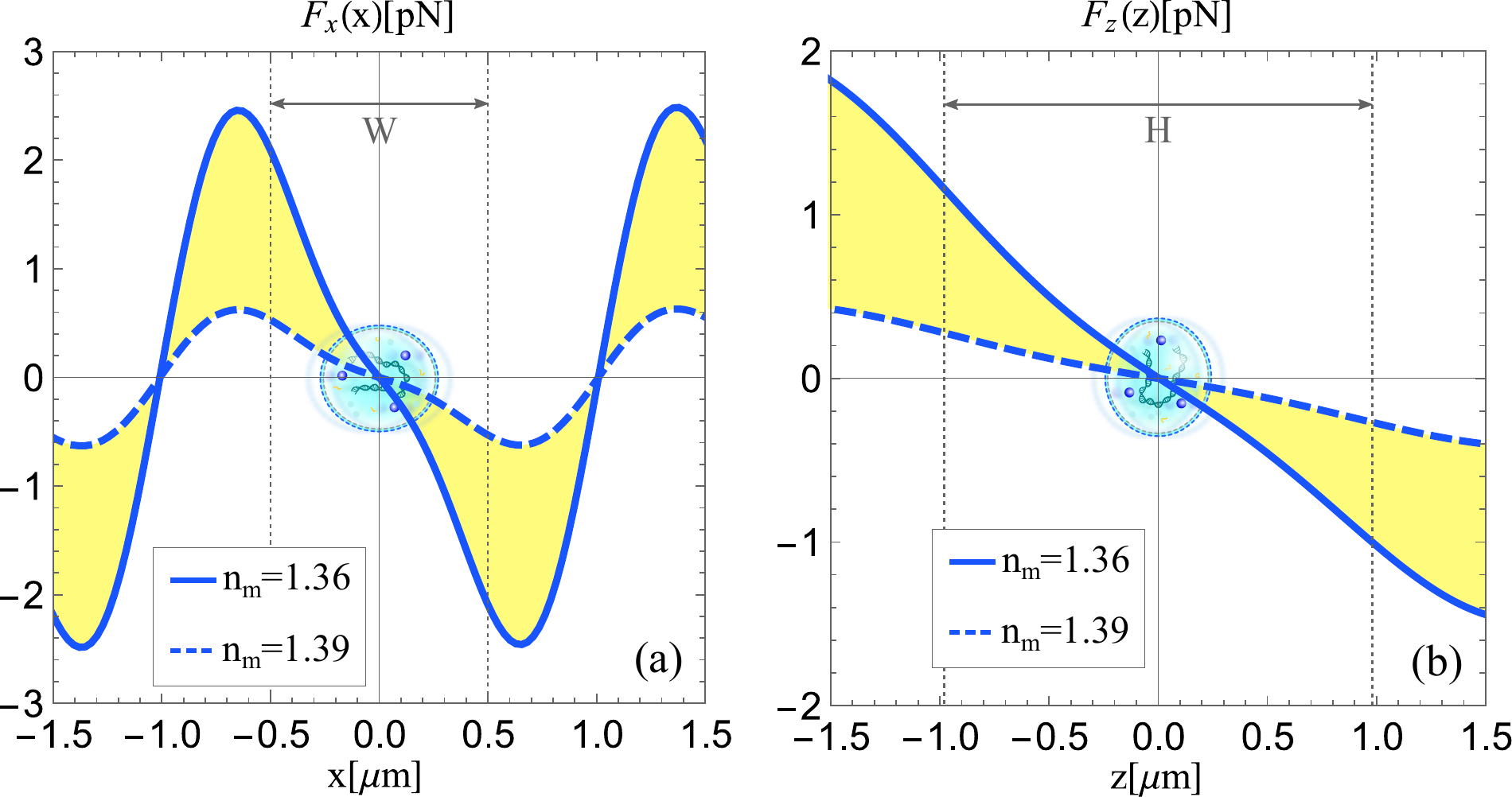}
    \caption{Forces acting on a trapped Mycoplasma cell, shown here centered at the origin for size comparison, (a) along the $x$ direction as a function of radial displacement and (b) along the $z$ direction as a function of axial displacement. The solid and the dashed curves correspond to a medium's refractive index of 1.36 and 1.39, respectively, while the yellow area correspond to $1.36\leq n_m\leq 1.39$. The other parameters used in the simulation are displayed in Table \ref{tab:exp}.}
    \label{fig:myco}
\end{figure}

Mycoplasma are known to be among the smallest living organisms, and perhaps the simplest cells \cite{Morowitz1984}. With radii around $\sim 0.3 \mu$m, these organisms lack a cell wall \cite{Krause2001}, being protected from the surrounding environment solely by their cellular membrane. This may present interesting mechanical and elastic properties which could be probed with the bottle beam. We propose investigating the trapping of Mycoplasma cells immersed in a non-toxic mixture of refractive index $~1.40$. Iodixanol presents a possible such medium, but further empirical tests must be carried over to fully determine how it affects living Mycoplasma cells. Figure \ref{fig:myco} shows the simulated forces acting on a trapped Mycoplasma when the parameters shown in Table \ref{tab:exp} are used. As it can be seen, forces along radial and axial directions are restorative and should provide stable trapping inside the bottle.

\begin{table}[h!]
\caption{Proposed values for trapping a Mycoplasma cell using a bottle beam. \label{tab:exp}}
\centering
\begin{tabular}{lcc}
\hline
\hline
Parameter  \hspace{2em}                     &          Units     \hspace{2em}      & Value             \\ \hline
Particle refractive index $n_p$       &       -                              & 1.36-1.39 \\
Medium refractive index $n_m$                  &     -                             & 1.40              \\
Particle radius $R$ &   $\mu$m                                  & 0.3            \\
Laser wavelength $\lambda_0$   &   nm   & 1064\\
Numerical aperture NA           &      -         & 0.7 \\
Laser power                          &          mW     & 500          \\
Index $p$                &          -     & 1\\
\hline\hline
\end{tabular}
\end{table}

It is known that direct incidence of focused laser light onto living cells can affect their division and growth \cite{Zhang2008}. As an interesting application of the bottle beam one could observe the process of cell division without directly sending a focused beam onto the trapped particle. 
The following experiment could be performed: at each round of measurement, a cell undergoing division is trapped in the dark focus by a given laser power and the complete cycle of the division process is observed. The power is then incrementally increased in every round of the experiment with a new cell, until a threshold value is reached at which cell division is significantly affected or perhaps even precluded. With the proper tweezer calibration presented in the previous section, the threshold power provides information on the forces acting during the process of cell division.

\section{Conclusions}
We have theoretically analysed the optical forces acting on a particle of lower refractive index than its surrounding medium trapped in the dark focus of a bottle beam generated by the superposition of a Gaussian beam and a Laguerre-Gauss beam with $\ell=0$ and $p\neq0$. Because %laser wavelengths and 
the size of trapped particles commonly range from tens of nanometers \cite{Hansen2005} to several microns \cite{Alinezhad2019} we analysed such forces both for particles much smaller than and with dimensions comparable to the wavelength of the trapping beam.

In the case of small particles, the dipole approximation was applied, resulting in a number of distinguishing features of the investigated trap. Scattering was found to be null at the focus of the beam, eliminating imbalance between gradient and scattering forces \cite{Nieminen2008}. The optical potential, on the other hand, coupled the motion along the radial and axial directions. It was shown that these could be decoupled by using a sufficiently high trapping power. The approximated decoupled potential turns out to be quartic and quadratic in the radial and axial directions, respectively. 
To test the validity of the approximated potential, motion of a particle trapped by the exact potential in a viscous medium was simulated and the results were confronted with those expected from the approximation as a function of laser power. 
%This confirmed that increasing the trapping power can be used to decouple the movement along different directions when using bottle beams to trap Rayleigh particles. 
We have also shown that by superposing a third mode, motion along the axial and radial directions can be decoupled independently of the trap power.  
To guide future experimental realizations, a calibration method was proposed.

In the case of larger objects, for which the dipole approximation is not valid, the tools developed in \cite{Nieminen2007, Lenton2020} were used to calculate the optical forces. Equilibrium positions after the focus were found, indicating a trapping regime different form the one described in \cite{Gahagan1996}. The interplay between the numerical aperture and the sphere's radius were explored and led to the conclusion that there is an upper bound for both of these quantities when using bottle beams for optical trapping. These limitations were interpreted in terms of the size of the optical bottle in comparison to the size of the particle, and were found to be eased by choosing a medium with refractive index close to that of the particle.

Finally, the findings obtained through exploration of the intermediate regime led to an experimental proposal to trap a living organism using the bottle beam. Considering values of refractive index reported in the literature, it is expected that trapping of small cells such as the Mycoplasma immersed in a non toxic high-refractive index medium in a dark focus is within reach.
This could be applied to situations in which focusing a high laser power onto the scrutinized cell is detrimental \cite{BlazquezCastro2019}, as in the case of cellular division \cite{Zhang2008}.

%\onecolumngrid
%\pagebreak
%\begin{center}
%\textbf{\large Appendix}
%\end{center}

%\setcounter{equation}{0}
%\setcounter{figure}{0}
%\setcounter{table}{0}

%\makeatletter
%\renewcommand{\theequation}{S\arabic{equation}}
%\renewcommand{\thetable}{S\arabic{table}}
%\renewcommand{\thefigure}{S\arabic{figure}}

\section*{Acknowledgements}
We would like to acknowledge Lucianno Defaveri for fruitful discussions regarding the statistical mechanics aspects of this work. 

In 2019, T.G. attended the Prospects in Theoretical Physics program at the Institute for Advanced Studies in Princeton. The meeting was centered on “Great Problems in Biology for Physicists” and had an important impact in the development of this work. 

This study was financed in part by the Coordena\c c\~{a}o de Aperfei\c coamento de Pessoal de N\'{i}vel Superior - Brasil (CAPES) - Finance Code 001, Conselho Nacional de Desenvolvimento Cient\'{i}fico e Tecnol\'{o}gico (CNPq), Funda\c c\~{a}o de Amparo \`a Pesquisa do Estado do Rio de Janeiro (Faperj, Scholarships No. E-26/200.270/2020 and E-26/202.830/2019), Instituto Serrapilheira (Serra-1709-21072) and Instituto Nacional 
de Ci\^encia e Tecnologia de Informa\c c\~ao Qu\^antica (INCT-CNPq).

\section*{Appendix A: total power of a bottle beam}
Throughout Section \ref{sec:dipole}, we used the power $P_0$ of each beam in the superposition as a measure of the trapping power. To find the exact relation between $P_0$ and the total power of the beam, we need to integrate equation \ref{eq:exact_intensity} along some plane orthogonal to the propagation of the beam. Choosing the plane $z=0$,
\begin{eqnarray}
        P&=&\int_{0}^\infty d\rho  \int_{0}^{2\pi}(d\theta\rho)I_0e^{-2\rho^2/\omega_0^2}\nonumber\\&\times&\left[1-2L^0_p\left(\frac{2\rho^2}{\omega_0^2}\right)+L^0_p\left(\frac{2\rho^2}{\omega_0^2}\right)^2 \right]\nonumber\\&=&2\pi I_0\int_{0}^\infty \frac{\omega_0^2du}{4} e^{-u}[1-2L_p^0(u)+L_p^0(u)^2]\nonumber\\&=&P_0\int_{0}^\infty du\,\, e^{-u}[1-2L_p^0(u)+L_p^0(u)^2]\label{eq:Ptotal}.
\end{eqnarray}

The Laguerre-Polynomials satisfy
\begin{equation}
    \int_0^\infty x^\ell e^{-x} L_p^{\vert\ell\vert}(x)L_q^{\vert\ell\vert}(x)=\frac{(p+\ell)!}{p!}\delta_{p,q},
\end{equation}
which implies that
\begin{eqnarray}
\label{eq:int1}
    \int_0^\infty du\,e^{-u}&=&\int_0^\infty du\,e^{-u}L^0_0(u)L^0_0(u)=1\\
    \label{eq:int2}
    \int_0^\infty du\,e^{-u}L_p^0(u)&=&\int_0^\infty du\,e^{-u}L^0_0(u)L^0_p(u)=0\\
    \label{eq:int3}
    \int_0^\infty du\,e^{-u}L_p^0(u)^2&=&\int_0^\infty du\,e^{-u}L^0_p(u)L^0_p(u)=\hspace{-1mm}1.
\end{eqnarray}
Finally, substituting Eqs. (\ref{eq:int1})-(\ref{eq:int3}) into Eq.(\ref{eq:Ptotal}), we find
\begin{equation}
   P=2P_0. 
\end{equation}

\section*{Appendix B: The decoupling mode}
In this section we derive the conditions that must be fulfilled by a three-mode superposition in order to provide a 
decoupled potential for the trapped particles, while keeping the bottle beam structure. 
Let us consider the superposition
\begin{equation}
\label{superp}
E (\rho, z) = E^{LG}_{0,0}(\rho, z) + \alpha_1 E^{LG}_{0,p_1}(\rho, z) + \alpha_2 E^{LG}_{0,p_2}(\rho, z)\,.
%\label{eq:3-mode}
\end{equation}
We want to obtain a relationship between the complex coefficients $\alpha_1$ and $\alpha_2\,$, and the radial orders 
$p_1$ and $p_2$ that achieve the desired decoupling and bottle-beam profile. The LG modes with zero OAM are given by
\begin{eqnarray}\label{LG}
\!\!E^{LG}_{0, p}(\bar{\rho}, \bar{z}) &=&\!\frac{\mathcal{N}}{\sqrt{1 + \bar{z}^2}} e^{- \frac{\bar{\rho}^2}{2}} L^{0}_p \left( \bar{\rho}^2 \right) 
\left(\frac{1-i\bar{z}}{\sqrt{1+\bar{z}^2}}\right)^{2p+1}\nonumber\\
&\times&\exp\left[ikz+ik\frac{\rho^2}{2R(z)}\right],
%\label{eq:LG-mode}
\end{eqnarray}
where we defined $\mathcal{N}=\sqrt{4P_0/c\epsilon\pi\omega_0^2}$, $\bar{\rho}=\sqrt{2}\rho/w(z)\,$, $\bar{z}=z/z_R$ and the last term is the Gouy phase.

The trapping potential is proportional to the light intensity distribution and, therefore, to the square modulus of 
the electric field
\begin{eqnarray}
\label{intensity}
I(\bar{\rho}, \bar{z}) &=& \frac{I_0}{1 + \bar{z}^2} \,e^{-\bar{\rho}^2}
\Bigg|1 + \alpha_1 L^{0}_{p_1} \left( \bar{\rho}^2 \right) \left(\frac{1-i\bar{z}}{\sqrt{1+\bar{z}^2}}\right)^{2p_1}\nonumber\\&+&\alpha_2 L^{0}_{p_2} \left( \bar{\rho}^2 \right) \left(\frac{1-i\bar{z}}{\sqrt{1+\bar{z}^2}}\right)^{2p_2}\Bigg|^2.
\end{eqnarray}
We seek an approximate expression for the trapping potential around the beam focus, which can be 
obtained from a power series expansion around this point. 
The bottle-beam condition requires that the light intensity vanishes at the focus. 
Note that $L^{0}_{p} (0) = 1\,$, so the light intensity at the beam focus is proportional to
\begin{equation}
\label{intensity0}
I(\mathbf{0}) = I_0
\big|1 + \alpha_1 + \alpha_2\big|^2,
\end{equation}
which implies
\begin{equation}
\label{bb-condition}
\big|1 + \alpha_1 + \alpha_2\big|^2 = 0\,.
\end{equation}
This condition cancels out the zero order contribution to the power series expansion. 
We will keep terms up to $\bar{\rho}^4$ and $\bar{z}^2\,$, which are the first non vanishing 
contributions to the power series. Since the zero order term vanishes, it will be easier to expand 
first the expression inside the square modulus in Eq. \eqref{intensity} and keep terms up to 
$\bar{\rho}^2$ and $\bar{z}^2\,$. The following approximations are assumed
\begin{eqnarray}
\label{approx}
e^{-\bar{\rho}^2} &\approx& 1 - \bar{\rho}^2\,,
\\
L^{0}_{p} \left( \bar{\rho}^2 \right) &\approx& 1 - p\bar{\rho}^2\,,
\\
\left(\frac{1-i\bar{z}}{\sqrt{1+\bar{z}^2}}\right)^{2p} &\approx& 1 - 2ip\bar{z} -2p^2 \bar{z}^2
\,.\\
\frac{1}{1+\bar{z}^2}&\approx& 1 - \bar{z}^2
\end{eqnarray}
Applying the approximations above together with the bottle-beam condition \eqref{bb-condition}, 
we find the following approximate expression for the trapping intensity
\begin{eqnarray}
\label{approx-intensity}
I(\Bar{\rho}, \Bar{z}) &\approx& 
I_0 \left(1-\bar{\rho}^2\right)
\big|2 A \bar{z} (\bar{z}\!-\!i\bar{\rho}^2)\!+\!B (\bar{\rho}^2 + 2i\bar{z})\big|^2
\nonumber\\
&\approx& I_0
\left[
|B|^2 (4\bar{z}^2\!+\!\bar{\rho}^4)-4 
\{|B|^2\!+\!\mathrm{Re}\left(A B^*\right)\}
\bar{\rho}^2 \bar{z}^2
\right]
\nonumber\\
&\approx& I_0\!\!\left[
4|B|^2\!\left(\frac{z^2}{z_R^2} \!+\!\frac{\rho^4}{w_0^4}\right)\!\!-8\!\left[|B|^2\!+\!\!\mathrm{Re}\left(AB^*\right)\right]\!\frac{z^2\!\rho^2}{z_R^2\!w_0^2}\right] ,
\nonumber
\end{eqnarray}
where we defined
\begin{eqnarray}
\label{defs}
A &=& \alpha_1 p_1^2 + \alpha_2 p_2^2\,,
\\
B &=& \alpha_1 p_1 + \alpha_2 p_2
\,.
%\label{eq:3-mode}
\end{eqnarray}
The two-mode bottle-beam potential is recovered by making $\alpha_1 = -1$ and $\alpha_2 = 0\,$.

Decoupling between the radial ($\rho$) and longitudinal ($z$) dependencies is achieved by choosing $\alpha_1$ and 
$\alpha_2$ such that 
\begin{eqnarray}
\label{decouple-condition}
|B|^2 + \mathrm{Re}\left(AB^*\right)=0\qquad (B\neq 0)\,.
\end{eqnarray}
We can write this condition in terms of the real and imaginary parts of the coefficients $\alpha_j = a_j + ib_j\,$. 
Including the bottle-beam condition, the following equations must hold
\begin{eqnarray}
\label{cond1}
&&1 + a_1 + a_2 = 0\,,\\
\label{cond2}
&&b_1 + b_2 = 0\,,\\
\label{cond3}
&&(a_1^2 + b_1^2)(p_1^2 + p_1^3) + (a_2^2 + b_2^2)(p_2^2 + p_2^3)
\nonumber\\
&& + p_1 p_2 (p_1 + p_2 +2)(a_1 a_2 + b_1 b_2) = 0\,.
\end{eqnarray}
By using \eqref{cond1} and \eqref{cond2} in \eqref{cond3}, we derive the following condition:
\begin{eqnarray}
\label{condfinal}
&&(a_1^2 + b_1^2) \left[p_1^2 (p_1+1) + p_2^2 (p_2+1) - p_1 p_2 (p_1 + p_2 + 2)\right] 
\nonumber \\
&&+a_1 \left[2 p_2^2 (p_2+1) - p_1 p_2 (p_1 + p_2 + 2)\right] + p_2^2 (p_2 + 1) = 0\,.
\nonumber\\
\end{eqnarray}
For example, let us set $p_1 = 1$ and $p_2 = 2\,$, giving
\begin{eqnarray}
\label{cond-example}
\!\!\!\!\!\!\!\!\!\!
a_1^2\!+\!b_1^2\!+\!\frac{7}{2}a_1\!+\!3 = 0\;\Rightarrow\; 
b_1^2 = -\!\left(\!a_1^2\!+\!\frac{7}{2} a_1\!+\!3\right)\!\geq\!0
\,.
\end{eqnarray}
This condition has infinite solutions in the interval $-2\leq a_1\leq -3/2\,$. Its limits provide real solutions 
for the superposition coefficients:
\begin{itemize}
\item $\alpha_1 = -2\;, \alpha_2 = 1\,$. 

\item $\alpha_1 = -3/2\;, \alpha_2 = 1/2\,$. 
\end{itemize}
Note that the first real solution is useless, since it gives $B=0$ and cancels out all terms up to $\rho^4$ and $z^2$ in the 
trapping potential. The other real solution gives $A=1/2$ and $B=-1/2\,$, resulting in the following expression for intensity of 
the electric field
\begin{eqnarray}
\label{decouple-intensity}
I(\rho, z) &\approx& 
I_0
\left(\frac{z^2}{z_R^2} + \frac{\rho^4}{w_0^4}\right)
\,,
\end{eqnarray}
which provides the desired bottle-beam configuration with decoupled dynamics along the transverse and longitudinal directions. Moreover, we can easily show that this solution is optimal. Under the bottle-beam and decoupling condition, the trapping 
strength is 
\begin{eqnarray}
\label{Bfinal}
4\vert B\vert ^2 = 2a_1 + 4\,,
\end{eqnarray}
which is a linear function of $a_1$ with positive slope. Therefore, its maximum value is obtained at the upper limit $a_1=-3/2\,$.

\bibliography{main.bib}

\end{document}